\documentclass[english,xaps,prb,twocolumn]{revtex4}
\usepackage{amsmath}
\usepackage{amssymb}
\usepackage{graphicx,color}

\makeatletter
\@ifundefined{textcolor}{}
{%
 \definecolor{BLACK}{gray}{0}
 \definecolor{WHITE}{gray}{1}
 \definecolor{RED}{rgb}{1,0,0}
 \definecolor{GREEN}{rgb}{0,1,0}
 \definecolor{BLUE}{rgb}{0,0,1}
 \definecolor{CYAN}{cmyk}{1,0,0,0}
 \definecolor{MAGENTA}{cmyk}{0,1,0,0}
 \definecolor{YELLOW}{cmyk}{0,0,1,0}
}

 \usepackage{amsbsy}
\usepackage{mciteplus}
\usepackage{babel}

\makeatother

\usepackage{babel}
\begin{document}

\title{Tight-binding simulations of electrically driven spin-valley transitions in carbon nanotube quantum dots}

\author{E. N. Osika}
\affiliation{AGH University of Science and Technology, Faculty of Physics and Applied Computer Science,\\
al. Mickiewicza 30, 30-059 Krak\'ow, Poland}
\author{A. Mre\'nca}
\affiliation{AGH University of Science and Technology, Faculty of Physics and Applied Computer Science,\\
al. Mickiewicza 30, 30-059 Krak\'ow, Poland}
\author{B. Szafran}
\affiliation{AGH University of Science and Technology, Faculty of Physics and Applied Computer Science,\\
al. Mickiewicza 30, 30-059 Krak\'ow, Poland}

\date{\today}

\begin{abstract}
We describe dynamics of spin and valley transitions driven by alternating electric fields
in quantum dots defined electrostatically within semiconducting carbon nanotubes (CNT). We use the tight-binding approach to describe
the states localized within a quantum dot taking into account the
circumferential spin-orbit interaction due to the s-p hybridization and external fields.
The basis of eigenstates localized in the quantum dot is used in the solution of the time-dependent Schr\"odinger equation
for description of spin flips and inter-valley transitions that are driven by periodic perturbation in the presence of coupling between the spin, valley and orbital degrees of freedom.
Besides the first order transitions we find also fractional resonances.
 We discuss the transition rates with selection rules that are lifted by atomic disorder and the bend of the tube.
We demonstrate that the electric field component perpendicular to the axis of the CNT activates spin transitions
which are otherwise absent and that the resonant spin-flip time scales with the inverse of the electric field.
\end{abstract}

\maketitle

\section{Introduction}

Electron spins confined in carbon nanotube\cite{cnt} (CNT) quantum dots\cite{cntqd} (QD) are considered attractive for quantum information storage and processing
due to the absence of the hyperfine interaction\cite{c13} which is the main source of decoherence in III-V nanostructures.
The spin-orbit (SO) coupling that is intrinsically present in CNTs due to s-p hybridization accompanying
the curvature of the graphene plane \cite{Ku,soc1,soc2,bulaev,soc3,klino1} paves the way for electrical control of the confined carrier spins.
In particular the SO interaction  allows for spin flips induced by AC electric fields  \cite{taconascytuja}
according to the mechanism of the electric-dipole spin resonance as studied earlier for III-V quantum dots.\cite{351,352,354,355,356}

In nanotube quantum dots the SO coupling splits the four-fold degeneracy
of energy levels with respect to the spin and valley
into Kramers doublets with spin-orbit coupling energy varying
from a fraction of meV \cite{Ku,jespersen} to several meV. \cite{large}
In this work we study the states confined in a QD defined electrostatically
within the CNT and simulate spin and valley transitions driven by AC electric field
between the quadruple of nearly degenerate energy levels in external magnetic field.

For clean CNTs the coupling between the $K$ and $K'$ valleys is absent which
motivates ideas to use the valley degree of freedom as a carrier of the quantum information alternative for the electron spin.
In the transport experiments the valley filters and valves were proposed \cite{Rycerz07} for clean samples in which the inter-valley scattering can be neglected. For clean CNT double quantum dots
the phenomenon of valley blockade has been demonstrated in experiment  \cite{ffpei}
and studied theoretically \cite{pal}
as the equivalent of the Pauli spin blockade. \cite{pcnd}
A theory for Rabi inter-valley resonance for CNT has also been presented \cite{palprl} within a continuum approximation
of the tight-binding Hamiltonian.

In this work we report on time-dependent tight-binding simulations for the spin-valley transitions
driven by AC field. In the present model the electron confinement within the dot, the lattice disorder, and the spin-valley dynamics are monitored at the atomic scale.
We work with a direct solution of the time dependent Schr\"odinger equation which
allows us to resolve not only the Rabi oscillations corresponding to the first order transition
but also the fractional resonances in higher-order transitions observed in EDSR experiments on III-V \cite{frac} as well as CNT \cite{ffpei} QDs.

We discuss the effects driving the spin-flips with a particular focus on the electric field component
that is perpendicular to the axis of the CNT, and which is bound to appear in experimental setups
with CNTs deposited or suspended above the gates.\cite{Ku,large,ffpei}
We show that a very similar
dynamics of transitions is obtained for a bent CNT.
The bend of the nanotube for electric dipole spin resonance in nanotubes was previously proposed\cite{flensberg} but in the context of the electron motion along the bend
in the external magnetic field and the resulting variation of the  effective Zeeman splitting.  In the present system the motion of the electron is limited to the QD area
and has a secondary effect on the transitions, still the bend of the nanotube in external {\it electric} field lowers the symmetry
of the eigenstates which allows for the spin flips.
We discuss the consequences of the perpendicular electric field, disorder and the bend of the CNT for selection rules
and transition times.

\section {Model}
\begin{figure*}[htbp]
\begin{tabular}{ll}
a) & b )  \\
\includegraphics[scale=0.32]{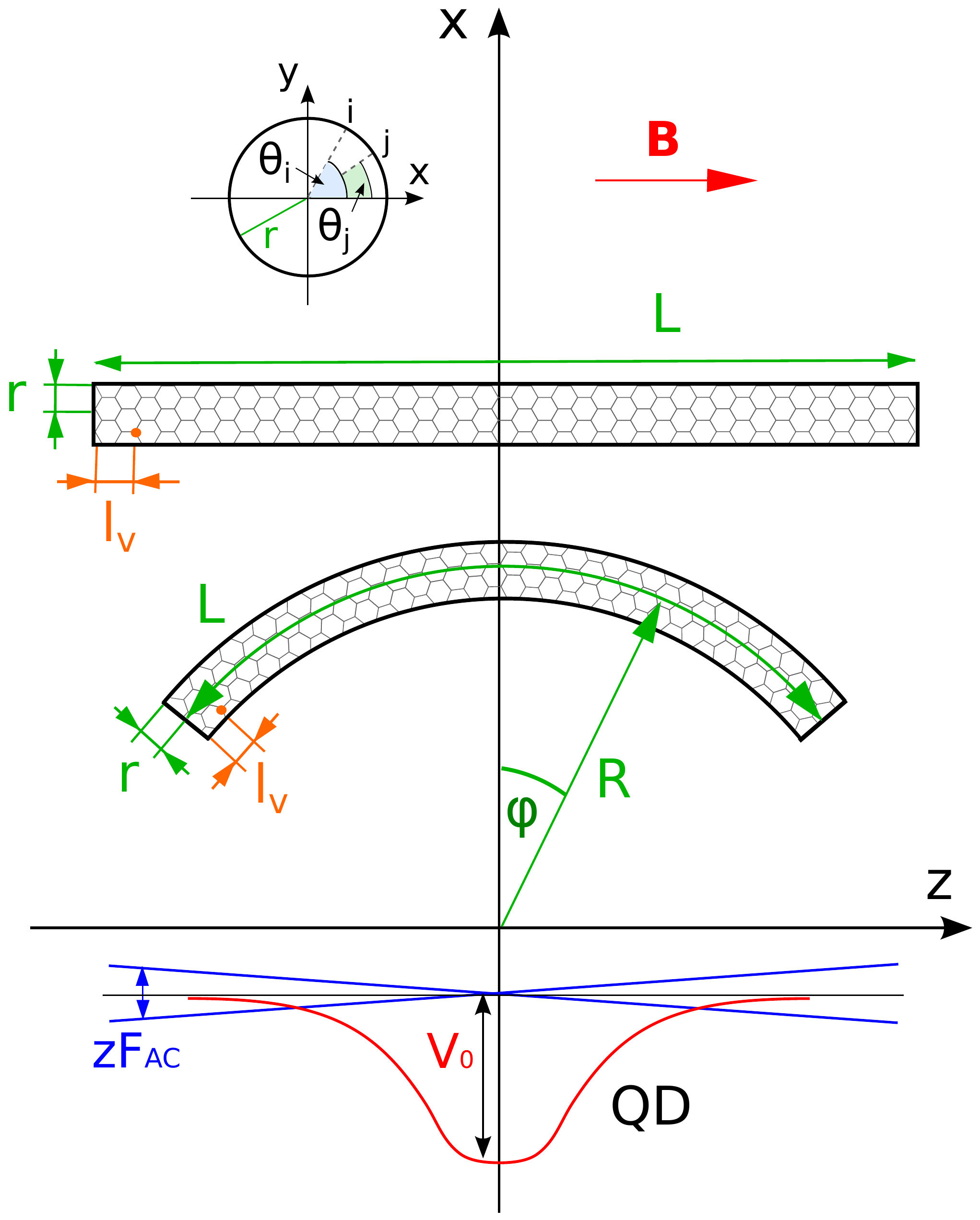} & \includegraphics[scale=0.3]{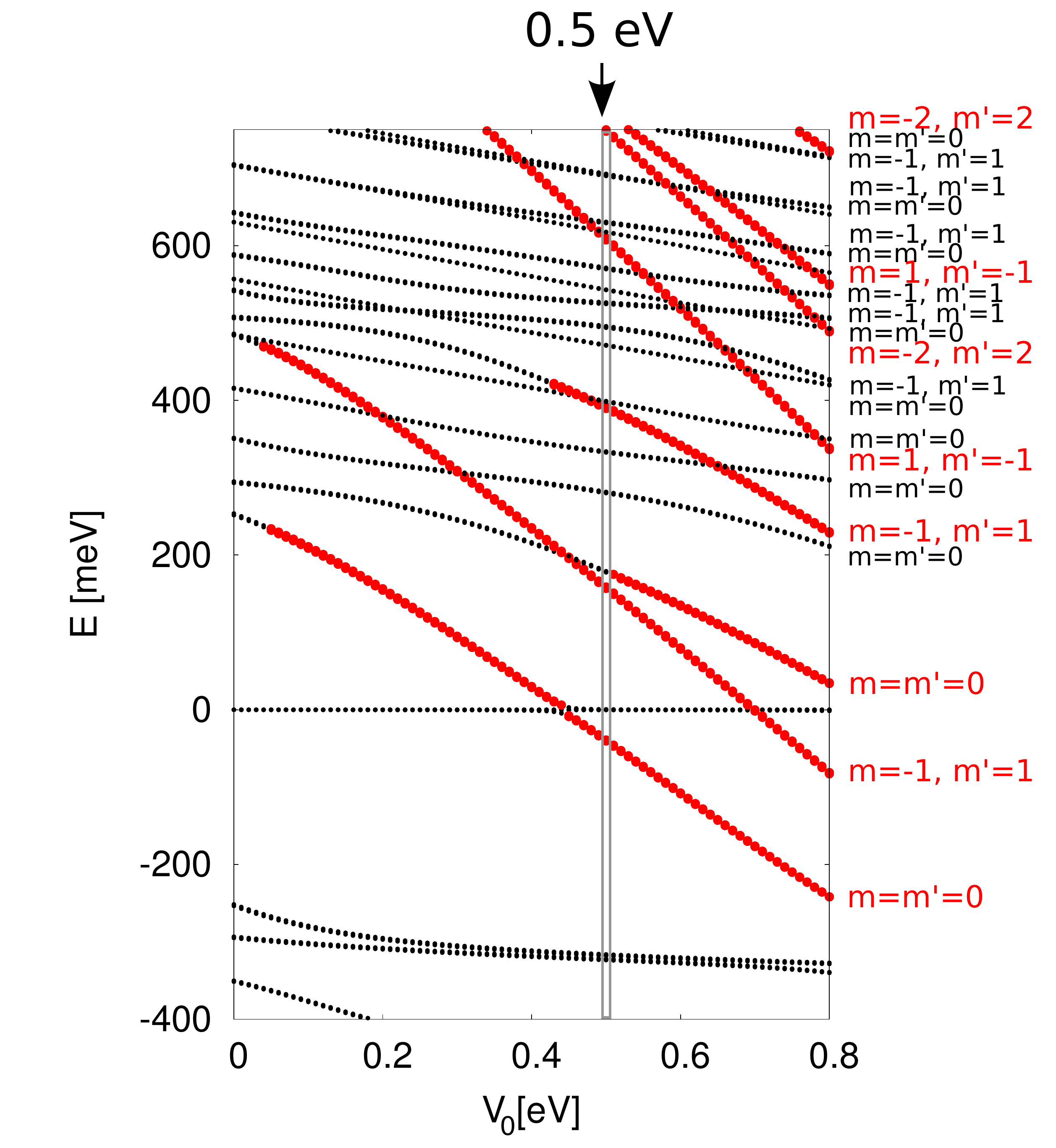} \end{tabular}
\caption{(a) Schematics of the considered system. In the top inset the definition of localization angles
of carbon atoms at a cross section of the tube perpendicular to its axis is given. Below: a straight and bent CNT
of length $L$ and  arc radius $R$. $r$ is the radius of the nanotube and $l_v$ is the distance of a vacant carbon atom removed in a part
of calculations from the end of the tube.
The bottom plot: the external confinement potential defining the QD and the potential due to AC field.
(b) Energy spectrum for the finite nanotube in function of the depth of the Gaussian confinement potential. With the red dots
we marked the energy levels with 50\% of the probability density within the QD area ($[-d,d]$). On this scale of the plot
the energy spectrum is indistinguishable for straight/bent CNTs, with or without SO coupling, $F_x=0$ or 100 kV/cm. 
On the right hand side of the figure the quantum numbers $m$ ($m'$) defining variation of the wave functions along the circumference
of the nanotube for the states of valley $K$ ($K'$).
}\label{schemat}
\end{figure*}

We consider a QD defined within a CNT by an external potential  that is depicted in Fig. \ref{schemat}(a).
The nanotube has a finite length $L=21.16$ nm with a diameter of $2r=1.56$ nm (after Ref. \onlinecite{jang}).
The CNT is placed symmetrically within $(x,z)$ plane with the external magnetic field applied along the $z$ axis.
We model the external potential, induced e.g. by a gate electrode, of a Gaussian \cite{gaussian} form
$W_{QD}({\bf r})=-V_0 \exp(-z^2/d^2)$, with the length of the quantum dot along the CNT given by $2d=0.2L$. In most of the simulations we take $V_0=0.5$ eV.
We consider both straight and bent CNTs [see Fig. \ref{schemat}(a)]. For the latter
we assume that the tube forms an arc of radius   $R=20$ nm [see Fig. 1(a)].
In order to simulate the atomic disorder within the tube we assume that one carbon atom is missing
within the lattice.  In Fig. 1(a) $l_v$ is the distance of the vacancy from the edge of the CNT.
Whenever the vacancy defect is introduced we take $l_v=0.85$ nm, i.e. fifth elementary cell from the edge of the tube.
In the calculations the external magnetic field is applied along the $z$ direction, unless stated otherwise.

{
In the present paper we choose to work with the tight-binding Hamiltonian instead of the continuum approximation.\cite{pal,palprl,bulaev,klino1,klino2}
The atomistic approach used here allows us to model directly the bend of the CNT and the effects of the atomic disorder 
which in the continuum version of the Hamiltonian needs to be introduced by a phenomenological intervalley scattering parameter. 
Moreover, the continuum approximation of the tight-binding Hamiltonian is limited to the low-energy part of the spectrum near the neutrality point.
The potential energy variation introduced by the external electric fields inducing the confinement is of the order of an electronvolt for which the 
low-energy approximation is hardly applicable. Note, that the low-energy continuum Hamiltonian for CNTs with spin-orbit coupling due to the curvature of the tube was derived in Ref. \onlinecite{klino1} from
the tight-binding model with application to the electric-dipole spin resonance. The low-energy theory was also applied for demonstration of helical modes in CNTs.\cite{klino2}}

We use the tight-binding Hamiltonian
\begin{eqnarray}
H&=&\sum_{\{i,j\}}(t_{ij} c_i^\dagger c_j+h.c.)\label{eqh}\\  \nonumber &+&\sum_i \left(W_i+\frac{1}{2}g\mu_b  \overrightarrow{\sigma}\cdot \overrightarrow{B}\right) c_i^\dagger c_i,
\end{eqnarray}
where the first summation runs over $2p_z$ spin-orbitals of nearest neighbor pair of atoms, $c_i^\dagger$ $(c_i)$ is the particle creation (annihilation) operator at ion $i$, and  $t_{ij}$ is the hopping parameter. The second sum in Eq. (\ref{eqh}) runs over all the $2p_z$ spin-orbitals where the term with the Land\`e factor $g=2$
introduces the spin Zeeman splitting with $\overrightarrow{\sigma}$ standing for the vector of Pauli matrices.
We found that inclusion of an external electric field perpendicular to the axis of the nanotube lifts the selection rules, in particular by
influencing the wave functions dependence on $\theta$ [Fig. 1(a)]. This component of the electric field appears naturally when
CNT is deposited above gates [cf. Refs. \onlinecite{Ku,large,ffpei}]. We account for the electric field potential $W_x=e F_x x$.
 The local potential at the $i$-th ion $W_i$ is taken as a superposition of the quantum dot potential and the potential
due to perpendicular external electric field $W_i=W_{QD}({\bf r}_i)+W_{x}({\bf r}_i)$.

In the tight-binding Hamiltonian the spin-orbit coupling is introduced via the spin dependence to the hopping parameters $t_{ij}$.\cite{soc1,chico}
The role of separate contributions to the spin-orbit coupling were discussed in detail in Ref. \onlinecite{soc2}.
The curvature induced coupling for the tube diameter of the order of 100 nm exceeds 3 times
the Rashba SO coupling produced by the external electric field of 1600 kV/cm, i.e.
of the order that is induced by the Gaussian potential assumed in this work.
The curvature induced SO coupling energy is inversely
proportional to the diameter of the nanotube.\cite{soc2} For the diameter of the nanotube considered here
we expect that the curvature-induced SO coupling is by two orders of magnitude stronger than the one due to external fields, so we neglect
the latter.
 The electric field that is perpendicular to the axis of the dot $F_x$ is taken equal to 100 kV/cm. The resulting
Rashba SO interaction is negligible.

The effects of CNT curvature including the spin-orbit coupling term are introduced
following Refs. \onlinecite{soc1} and  \onlinecite{soc3}, with the hopping terms
given by
\begin{eqnarray}
t_{\uparrow\uparrow}&=&\langle z_{i}\uparrow_{z}|H|z_{j}\uparrow_{z}\rangle=V_{pp}^{\pi}\cos(\theta_{i}-\theta_{j})\nonumber \\&-&(V_{pp}^{\sigma}-V_{pp}^{\pi})\frac{r^{2}}{a_{C}^{2}}[\cos(\theta_{i}-\theta_{j})-1]^{2}
\nonumber \\
&+&2i\delta\big\{ V_{pp}^{\pi}\sin(\theta_{i}-\theta_{j})+ \\&&(V_{pp}^{\sigma}-V_{pp}^{\pi})\frac{r^{2}}{a_{C}^{2}}\sin(\theta_{i}-\theta_{j})[1-\cos(\theta_{i}-\theta_{j})]\big\} \nonumber
\end{eqnarray}
\begin{eqnarray}
t_{\downarrow\downarrow}&=&\langle z_{i}\downarrow_{z}|H|z_{j}\downarrow_{z}\rangle=\langle z_{i}\uparrow_{z}|H|z_{j}\uparrow_{z}\rangle^{*}
\\
t_{\uparrow\downarrow}&=&\langle z_{i}\uparrow_{z}|H|z_{j}\downarrow_{z}\rangle= \\&&-\delta(e^{-i\theta_{j}}+e^{-i\theta_{i}})(V_{pp}^{\sigma}-V_{pp}^{\pi})\frac{rZ_{ji}}{a_{C}^{2}}[\cos(\theta_{i}-\theta_{j})-1] \nonumber
\\
t_{\downarrow\uparrow}&=&\langle z_{i}\downarrow_{z}|H|z_{j}\uparrow_{z}\rangle=-\langle z_{i}\uparrow_{z}|H|z_{j}\downarrow_{z}\rangle^{*},\label{spino}
\end{eqnarray}
where $z_{i}$ stands for $2p_{z}$ orbital on $i$-th atom, $\uparrow_{z}$ and  $\downarrow_{z}$ indicate the orientation of the spin along the $z$ direction,
 $V_{pp}^{\pi}=-2.66\,$eV, $V_{pp}^{\sigma}=6.38\,$eV,\cite{tomanek}
$a_C=0.142$ nm is the distance between the nearest neighbor atoms,
$\theta_{i}$ indicates the localization angle of atom $i$ in the $(x,y)$ plane [see the top inset in Fig. \ref{schemat}(a)],
and  $Z_{ji}=Z_{j}-Z_{i}$ is the distance between atoms $i$ and $j$ in the $z$ direction.
We adopt the SO coupling  parameter $\delta=0.003$ after Refs. \onlinecite{soc1} and \onlinecite{soc3}.

In the present model the spin-orbit interaction is uniquely due to the curvature
of the graphene plane. Since the bend of the entire nanotube [Fig. 1(a)] is weak as compared
to the graphene folding curvature ($R/r=20/0.78\simeq 25$) we do not take into account any extra spin-orbit interaction
due to a finite value of $R$, i.e. the values of $t_{ij}$ include the SO coupling due to the $\sigma$-$\pi$ hybridization that appears with the curvature of the graphene plane and are fixed for a {\it straight} CNT.
The finite value of $R$ is accounted for in the electric  $W_i$ and vector potential. 
As for the latter, in presence of the external magnetic field the hopping parameters acquire
an additional Peierls phase \begin{equation}
t_{nm}\rightarrow t_{nm}e^{i2\pi\Phi_{nm}}
\end{equation}
with the Aharonov-Bohm phase $\Phi_{nm}=(1/\Phi_{0})\int_{r_{n}}^{r_{m}}\boldsymbol{A}\cdot\boldsymbol{dl}$,
and the flux quantum $\Phi_{0}={h}/{e}$.

The folding of the CNT is defined by the chiral vector  $\boldsymbol{C_{h}}=n_{1}\boldsymbol{a_{1}}+n_{2}\boldsymbol{a_{2}}$,
where $\boldsymbol{a_{1}}=a_{0}(1,0)$, $\boldsymbol{a_{2}}=a_{0}(1/2,\sqrt{3}/2)$
and $a_{0}=0.246$ nm [see Fig. \ref{siec}].
In this paper  we assume a zigzag CNT with $n_2=0$, which possesses a semiconducting character with
the energy gap allowing for electrostatic confinement of the carriers provided that $n_1$ is not
a multiple of 3.\cite{cnt} In the following we take $n_1=20$ atoms along the circumference of the tube.

The spin-orbital eigenstates of Hamiltonian $H$, that are localized in the quantum dot induced by the external Gaussian potential are used for construction of the basis
in which the EDSR is simulated,
\begin{equation}
\Psi(\boldsymbol{r},\sigma,t)=\sum_{n}c_{n}(t) \Psi_{n}(\boldsymbol{r},\sigma)e^{-\frac{iE_{n}t}{\hbar}}. \label{basa}
\end{equation}
We consider the AC electric field that is applied along the nanotube,
which introduces time dependence to the Hamiltonian
\begin{equation}
H'(t)=H+eF_{0}z\sin(\omega t), \label{nh}
\end{equation}
with the amplitude of the oscillation taken equal to $F_{0}=4$ kV/cm and tunable frequency $\omega$.
The wave function (\ref{basa}) inserted to the time-dependent Schr\"odinger equation $i\hbar\frac{d\Psi}{dt}=H'\Psi$,
upon projection on the basis of dot-localized $H$ eigenstates gives a system of differential equations for the dependence of the
expansion coefficients on time
\begin{equation}
i\hbar c'_{k}(t)=\sum_{n}c_{n}(t)eF_0\sin(\omega t)\langle \Psi_{k}|z|\Psi_{n}\rangle e^{-\frac{i(E_{n}-E_{k})t}{\hbar}},
\end{equation}
which we solve for $c_k(t)$ using the implicit Crank-Nicolson scheme.

\begin{figure}[htbp]
\includegraphics[scale=0.32]{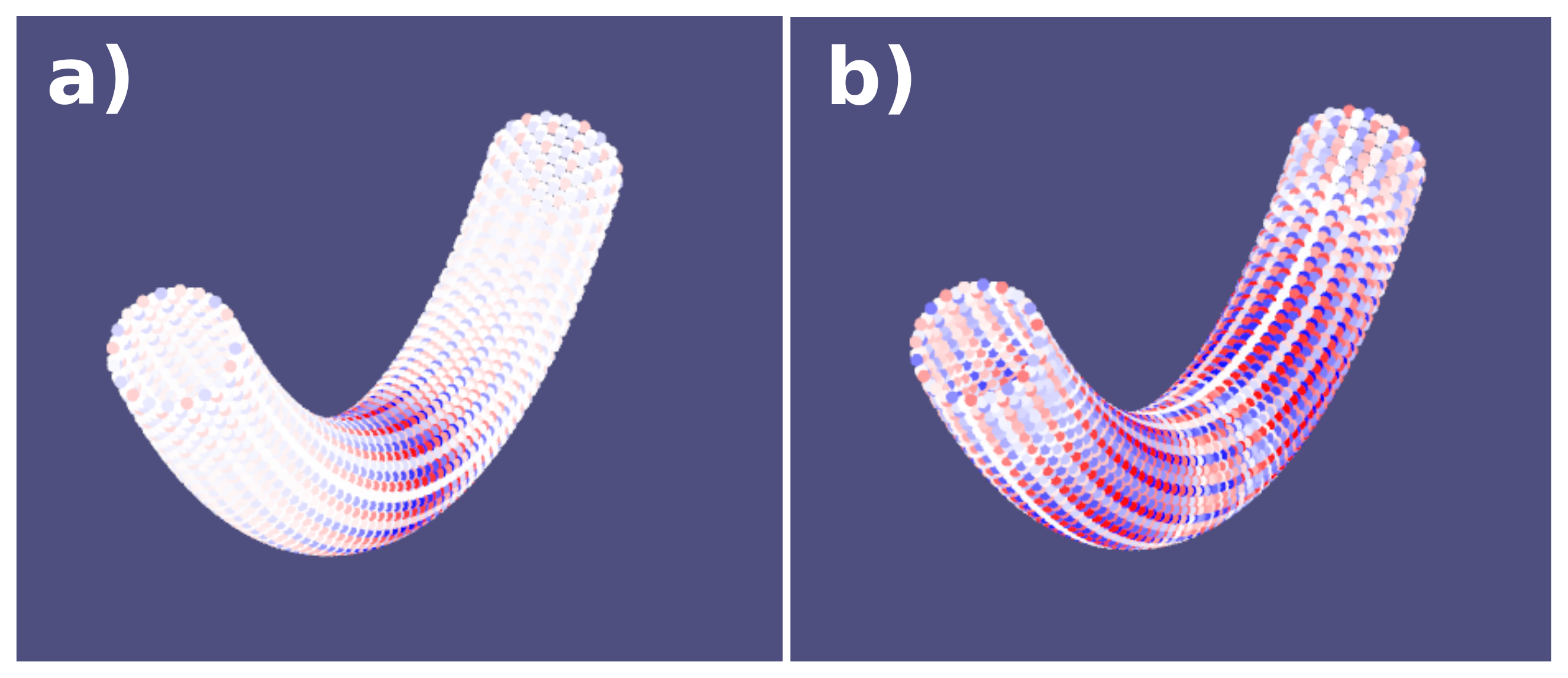}
\caption{
(a) The wave function of the lowest-energy QD localized level of Fig. \ref{schemat}(b) for $V_0=0.5$ eV of energy -40 meV.
Real part of the majority (spin-up) component of the $m'=0$ wave function corresponding to $K'$ valley.
(b) Real part of the majority (spin-up) component of the $m'=0$ wave function corresponding to $K'$ valley but delocalized
with the energy of 280 meV [see Fig. \ref{schemat}(b)]. The values are given at the carbon atoms
with red and blue colors corresponding to the opposite signs of the wave function.
}\label{mid}
\end{figure}

\begin{figure}[htbp]
\includegraphics[scale=0.72]{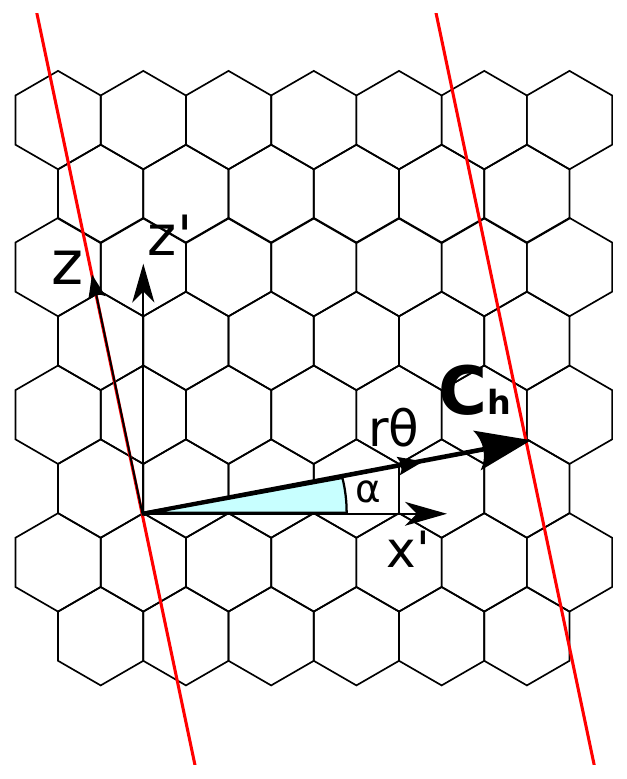}
\caption{
The chiral vector ${\bf C}_h$  defining the folding of the nanotube and system of the coordinates
used in the text. }\label{siec}
\end{figure}

\begin{figure}[htbp]
\includegraphics[scale=0.42]{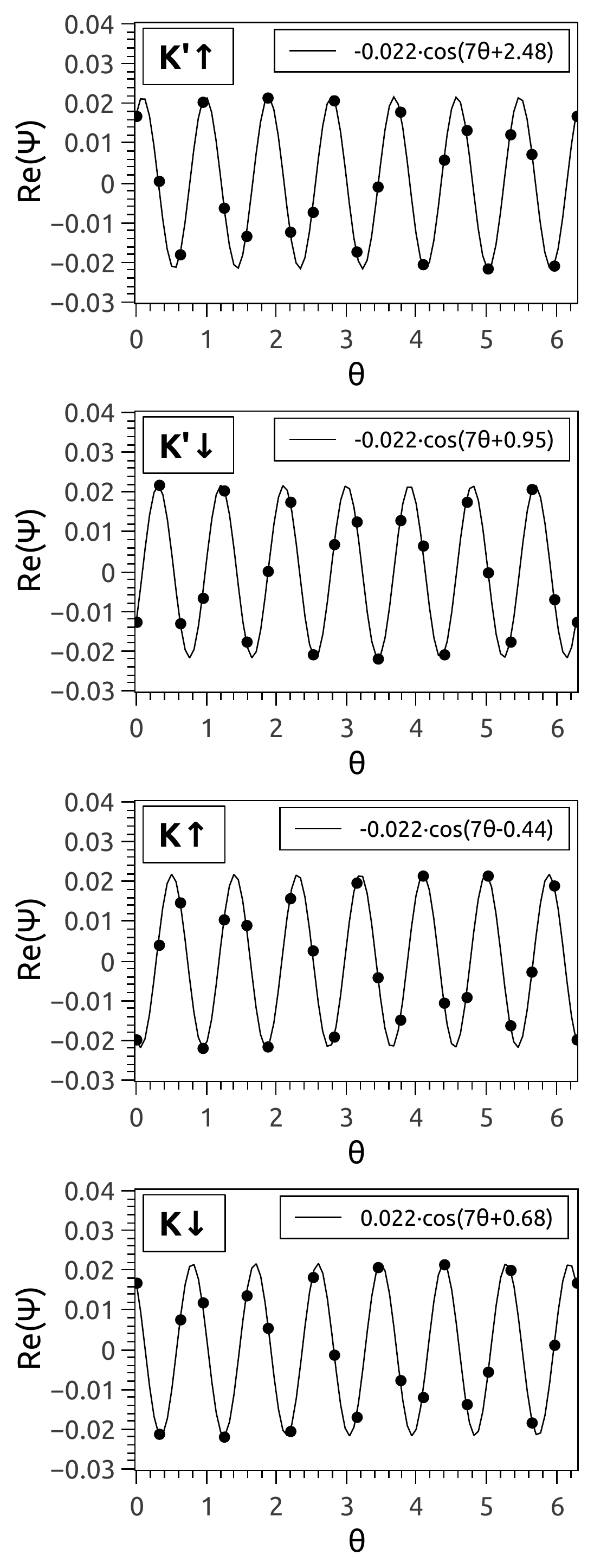}
\caption{Real part of the spin majority wave function component of the nearly degenerate lowest QD localized level of the energy $\simeq-40$ meV for $V_0=0.5$ eV.
The dots indicate the values of the wave functions at the 20 atoms along the CNT circumference and the solid line
indicates the wave function of the continuum approximation after Ref. \onlinecite{bulaev}.
} \label{dm}
\end{figure}

\begin{figure}[htbp]
\includegraphics[scale=0.45]{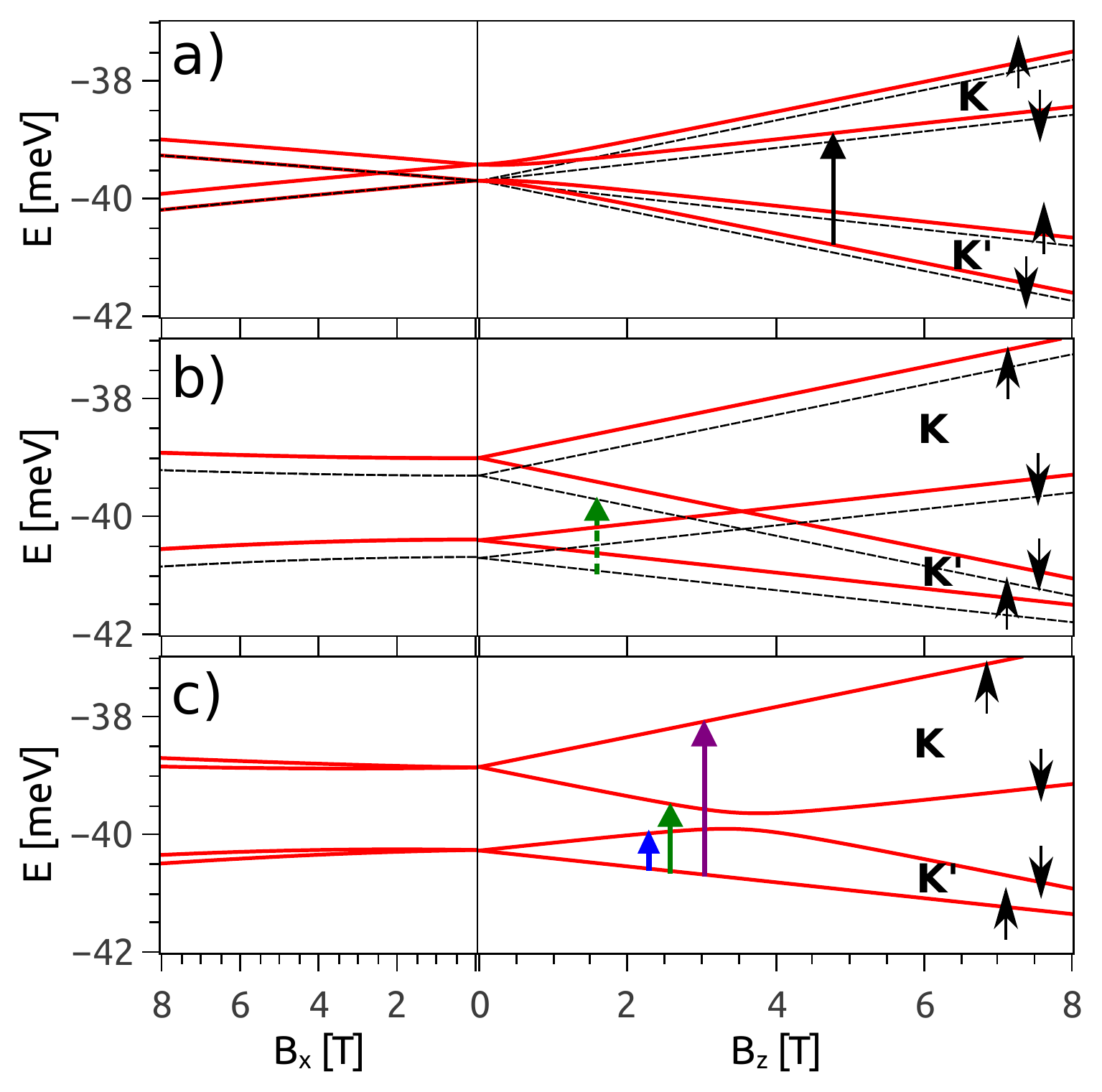}
\caption{Energy spectrum for the lowest-energy QD localized electron energy level for $V_0=0.5$ eV as a function of the
external magnetic field oriented along the $x$ or $z$ axis.  In (a)
a straight CNT is considered without SO coupling ($\delta=0$). The black energy levels correspond to a CNT without disorder.
The red energy levels were obtained for a vacant carbon atom (see Fig. \ref{schemat} for location of the defect).
(b) The energy spectrum with SO coupling ($\delta=0.003$) for a straight CNT (red lines) and for a bent one (black lines).
(c) Spectrum for a straight CNT with a vacancy [see Fig. 1(a)].
The vertical arrows in all the figures indicate the spin/valley transitions discussed in the text.}
\label{widmozb}
\end{figure}

\section{The stationary states}

Figure 1(b) shows the energy spectrum of states within the nanotube in function of $V_0$ in the absence of the
external magnetic field. At the scale of this figure, for straight / bent nanotube, with or without a component of the electric field in the $x$ direction, the spectrum is identical.
The red lines indicate the energy levels with more than 50\% of the probability density localized within the quantum dot.
The zero energy level that is independent of $V_0$ is related to states localized at the Fujita edge of the CNT. We verified
that the exact form of the edges at the end of the tube does not perturb the localized part of the spectrum.
We find that the angular dependence of the localized states very well agrees with the continuum theory of Ref. \onlinecite{bulaev}.
For nanotube with $(n_1,n_2)=(20,0)$ that we consider here
the angular dependence (fixed $z$) of the {\it majority} spin component
of the eigenstates of the continuum Hamiltonian is of the form  \cite{bulaev}
$\Psi_n=A \exp(i(m\pm 7)\theta)$,  where $m$ is the angular quantum number, with $+$ for the $K$ valley
and $-$ for $K'$ valley, and $A$ standing for a constant that is independent of $\theta$.
At the scale of  Fig. 1(b) all the energy levels appear as nearly four-fold
degenerate with respect to the valley and the spin.\cite{ff,ff1,ff2,ff3}

Figure \ref{dm} shows the angular dependence
of the real part of the lowest-energy electron QD confined states [the energy level marked in red with the energy of -40 meV for $V_0=0.5$ eV of Fig. 1(b)]
as obtained with the tight-binding approach (dots) and a fit with the wave function of the continuum approximation.\cite{bulaev}
We notice that all the states of these nearly degenerate energy levels correspond to $m=0$.
In a similar manner it is possible to identify the quantum numbers $m$ for all the tight-binding wave functions within
the nanotube, including the ones localized in the QD.
The subsequent quantum numbers are given in Fig. 1(b) with primes for the $K'$ valley.

Our identification of the valley is based on the energy dependence on the magnetic field along the CNT axis within the continuum approach
for a rectangular potential defined within the CNT,\cite{bulaev}
with  $E_{m,k_n}=\pm \sqrt{(m+\frac{\Phi_{AB}}{\Phi_{0}}+\frac{1}{3})^{2}/r^{2}+k_n^{2}}$ for $K$ valley (for $K'$ valley one needs to replace $m$ by $m'$ and $\frac{1}{3}$ by $-\frac{1}{3}$) where
$\Phi_{AB}=B\pi r^2$ and $k_n$ are the discrete wave vectors corresponding to localization of the wave function within the quantum dot.\cite{bulaev}
For $m=0$ and $B>0$ the $K'$ ($K$) valley energy levels go down (up) in the energy.

The value of $V_0=0.5$ eV is adopted for simulation of the spin and valley transitions. The work point is marked
with the vertical line in Fig. 1(b). We assume that the Fermi energy -- pinned by contacts -- is near the neutrality
 point ($E=0$). The CNT is overall charge neutral but with a single excess electron in the quantum dot occupying the nearly four-fold degenerate red energy level that is below
$E=0$ for $V_0=0.5$ eV [Fig. 1(b)]. Transitions between the quadruple of states (see below) within the red energy level are discussed below.
We assume that the potential of all the carbon ions and the electrons of the filled valence band is included in the tight-binding Hamiltonian parameters,
and that the valence band electrons do not react to the presence of the excess charge nor to the AC electric field. This is justified
by the energy gap between the lowest valence energy levels and the quantum-dot energy levels. The value of the gap -- about 220 meV is large
compared to the thermal excitation energy even at the room temperature.
In the time-dependent calculations we include to the basis [Eq. (7)] the unoccupied dot-localized energy levels [the red ones in Fig. 1(b)],
but the transitions -- for the adopted AC frequency occur only within the lowest -- nearly degenerate -- energy level.
The four states which form this energy level differ by the spin and the valley but the spatial charge distribution within
the system is almost identical. The Coulomb potential -- acting on the valence band electrons -- of the QD is unchanged when the transitions occur.
The valence band electrons will react to the external potential forming the QD which is large. The result of the reaction will be the screening of the external
potential. The potential that we use in this paper for the QD confinement should be considered as the effective one including the screening
by the valence band electrons.

The splitting of the lowest-energy electron level localized in the dot by the external magnetic
field as obtained for the present
model system is illustrated in Fig. \ref{widmozb}.
The panel (a)  of Fig. \ref{widmozb} shows the spectrum without SO coupling for
a straight CNT without disorder (black lines) and with the vacancy at 0.84 nm from the edge
of the tube (red lines).
For the magnetic field perpendicular to the axis of the dot one obtains energy level splitting by the spin Zeeman effect only.
The field along the axis splits also the valley degeneracy. The coupling of the valleys
introduced by the vacancy splits the energy level at $B=0$.
When the SO coupling is accounted for [Fig. \ref{widmozb}(b)]
the spectrum for $B=0$ is split to spin-valley doublets \cite{Ku} even without disorder.
The spin-valley doublets are only split by the field along the $z$ direction. The impact of the CNT bend [dark lines in Fig. \ref{widmozb}(b)]
to the spectrum is weak -- there is a small shift to lower energies -- mostly due to the fact
that the bend introduces more carbon atoms within the QD area.
When the vacancy is introduced [Fig. \ref{widmozb}(c)] the coupling of the valleys produces avoided
crossing between $K$ and $K'$ energy levels with spins oriented antiparallel to the magnetic field vector.
The vertical arrows in Fig. \ref{widmozb} indicate the spin-valley transitions which are observed
in the AC electric field and which are discussed below.
\begin{figure}[htbp]
\includegraphics[scale=0.45]{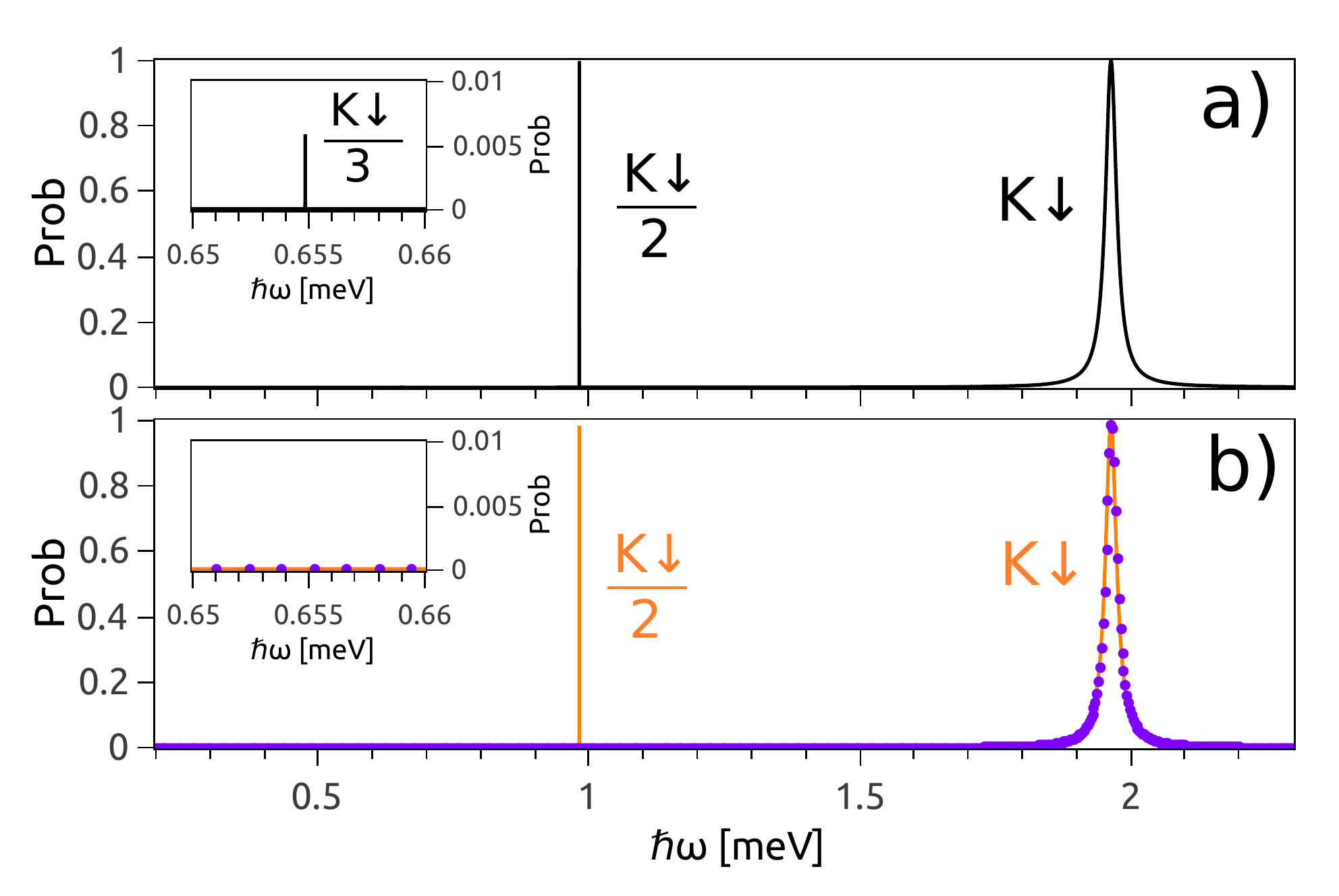}
\caption{(a) Results of the solution of the time-dependent Schr\"odinger equation for the straight CNT without SO coupling but with a vacancy [cf. red lines in Fig. \ref{widmozb}(a)] in the carbon lattice.
We plot the maximal occupation of the $K  \downarrow$ energy level for simulations lasting 500 ns starting from
the $K'\downarrow$ ground-state as a function of the AC driving frequency $\omega$. The wide and narrow peaks correspond to the direct and half-resonances.
The inset shows the zoom for the 1/3 of the excitation energy.
(b) Same as (a) only calculated with the the time dependent perturbation theory in the first (dots) and second (line) order see subsection \ref{tdpt}.}\label{ss}
\end{figure}

\begin{figure}[htbp]
\includegraphics[scale=0.4]{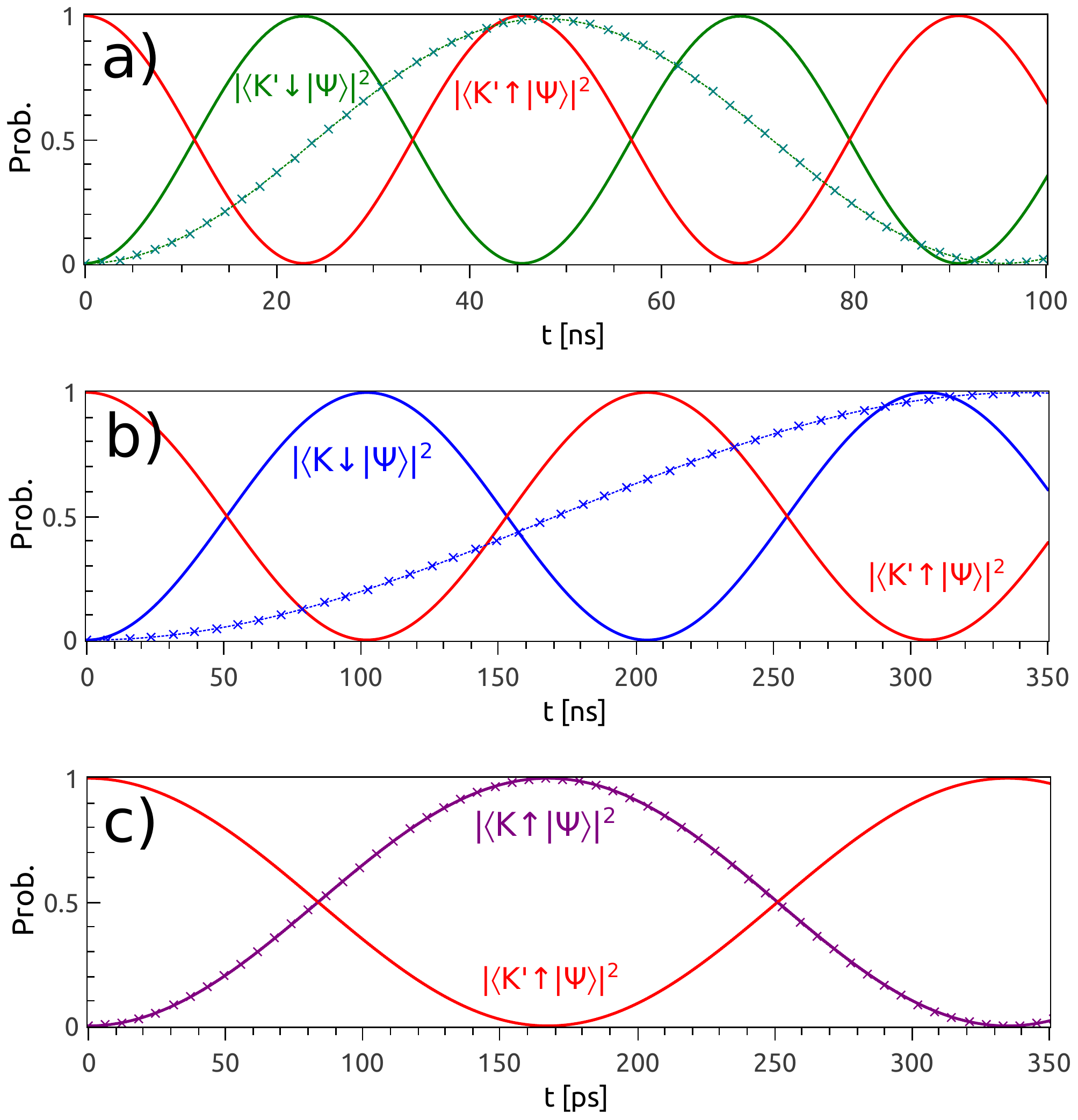}
\caption{ Time evolution for a straight CNT with SO coupling and a vacancy present within the system (the line
marked with crosses). The solid lines without crosses correspond to results with an additionally introduced
electric field of $F_x=100$ kV/cm in the direction perpendicular to the axis of the system. The resonant frequencies were set
for transitions from the ground state $K'\uparrow$  to $K'\downarrow$ (a) [spin flip -- see the green arrow in Fig. \ref{widmozb}(c)], $K\downarrow$ (b) [spin flip with inter-valley transition -- the blue arrow in Fig. \ref{widmozb}(c)], and $K\uparrow$ in (c) [inter-valley transition -- the purple arrow in Fig. \ref{widmozb}(c)].
The figures show the probabilities to find the system in the initial and final $f$ state
$|\langle \Psi(\boldsymbol{r},\sigma,t)|K'\uparrow\rangle|^2$ and $|\langle \Psi(\boldsymbol{r},\sigma,t)|f \rangle|^2$
of the resonant transition.}
\label{time}
\end{figure}

\begin{figure*}[htbp]
\includegraphics[scale=0.5]{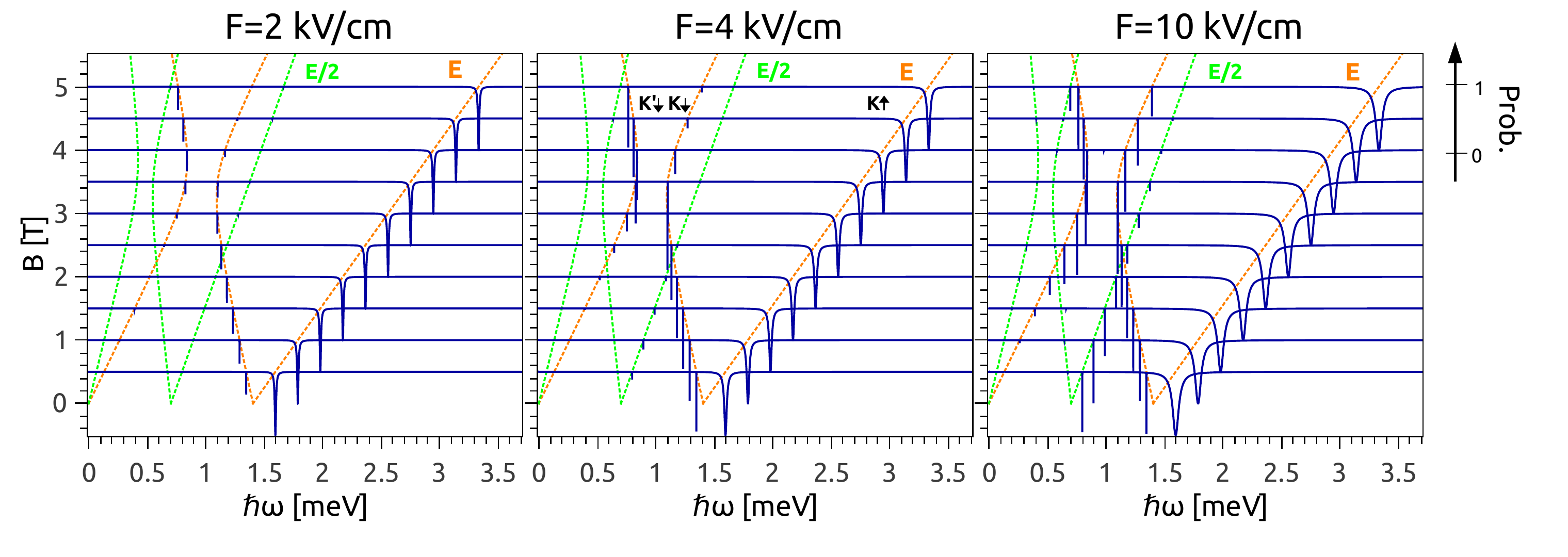}
\caption{The orange lines show the energy spectrum calculated with
respect to the ground state, i.e. the direct transition energies from the $K'\uparrow$ energy level.
The green ones show half of the direct transition energies.
The dark blue lines plot the minimal occupation probability of the ground state (initial one for the simulation) calculated with spacing of 0.5 T.
Each simulation lasted 20 ns.}
\label{szkant}
\end{figure*}

\section{Spin-valley transitions}

For simulations of the spin-valley transitions we set the external magnetic field
along the $z$ direction with  $B=5$ T. The time evolution in AC field is started from the lowest-energy state.
In presence of SO coupling [see Fig. \ref{widmozb}(b,c)] the ground state at $B_z=5$ T is the spin-up state of $K'$ valley.
Naturally with SO and valley-orbit coupling present the designates of valley and spin are approximate and not strict.
For higher magnetic field outside the Figure there is a spin transition in the ground state .

\subsection{CNT without SO coupling}
In the absence of SO coupling [see Fig. \ref{widmozb}(a)] the spin-down ground-state is promoted
by arbitrarily weak magnetic field. For the CNT without the vacancy [black lines in Fig. \ref{widmozb}(a)]
the system does not respond to AC field driving transition from the ground-state to any of the energy
levels given in the figure, which are of a different valley and/or opposite spin.
For the CNT including the vacancy [energy levels marked in red in Fig. \ref{widmozb}(a)] we do observe an inter-valley transition
with conserved spin. The scan of the driving frequencies for these valley transitions is given in Fig. \ref{ss}.
In Fig. \ref{ss} we plot the maximal value of $|\langle \Psi(\boldsymbol{r},\sigma,t)|K\downarrow\rangle|^2$ that is obtained
during 500 ns time evolution of the system.  Besides the direct (first-order) transition near $\hbar \omega=2$ meV, which
was previously studied by the continuum approximation to the CNT without SO coupling but with phenomenologically introduced disorder \cite{palprl}
we find the fractional one at half of the basic frequency for the direct transition.

\subsection{Straight CNT with SO coupling with disorder}
We introduced the SO coupling to the straight CNT still with a vacancy present and in Fig. \ref{time} we plotted the dynamics of transitions
from the ground-state $K'\uparrow$ to the three other energy levels of Fig. \ref{widmozb}(c).
For each of the plots: the valley transition [Fig. \ref{time}(c)], the spin flip [Fig. \ref{time}(a)]
and the transition including both valley and spin [Fig. \ref{time}(b)] we set the external frequency to the resonant one.
The lines with (without) the crosses in Fig. \ref{time}(b) were obtained in the absence (presence)
of the external field $F_x=100$ kV/cm.
This field has no detectable effect on the energy spectra of Fig. \ref{widmozb}(b,c).
Still its effect on transitions involving spin-flip [Fig. \ref{time}(a,b)] is distinct and it relies
on shortening the transition time two to three times. The electric field applied in the $x$ direction has no effect on the fast inter-valley transition [Fig. \ref{time}(c)],
which is by two orders of magnitude
shorter than the spin flip with preserved valley [Fig. \ref{time}(a)] and by three orders of
magnitude shorter than the transition involving change of both the valley and the spin [Fig. \ref{time}(c)].

Figure \ref{szkant} shows the transition energy spectrum and the transition effectiveness for the straight CNT
including the electric field in the $x$ direction for various amplitudes of the AC field.
Orange lines correspond to direct transitions and the green ones to resonances at half of the basic frequency for the direct transitions.
The duration of every simulation was set to 20 ns only. For the weakest amplitude ($F=2$ kV/cm)
we see the pronounced inter-valley transition with preserved spin [same as in Fig. \ref{time}(c)].
The other line that is observed is the spin-flip transition with preserved valley.
The line with both spin and valley transition is only detectable at the avoided crossing between
the valley states. For $F=4$ kV/cm the traces of the half resonant inter-valley transition appear.
For $F=10$ kV/cm the half-resonant transitions are intensified. Note that the half resonance
for low $B$ is more effective than the direct inter-valley transition with spin flip.
The half-resonant the inter-valley transition is less effective for higher $B$.

 \begin{figure}[htbp]
\includegraphics[scale=0.32]{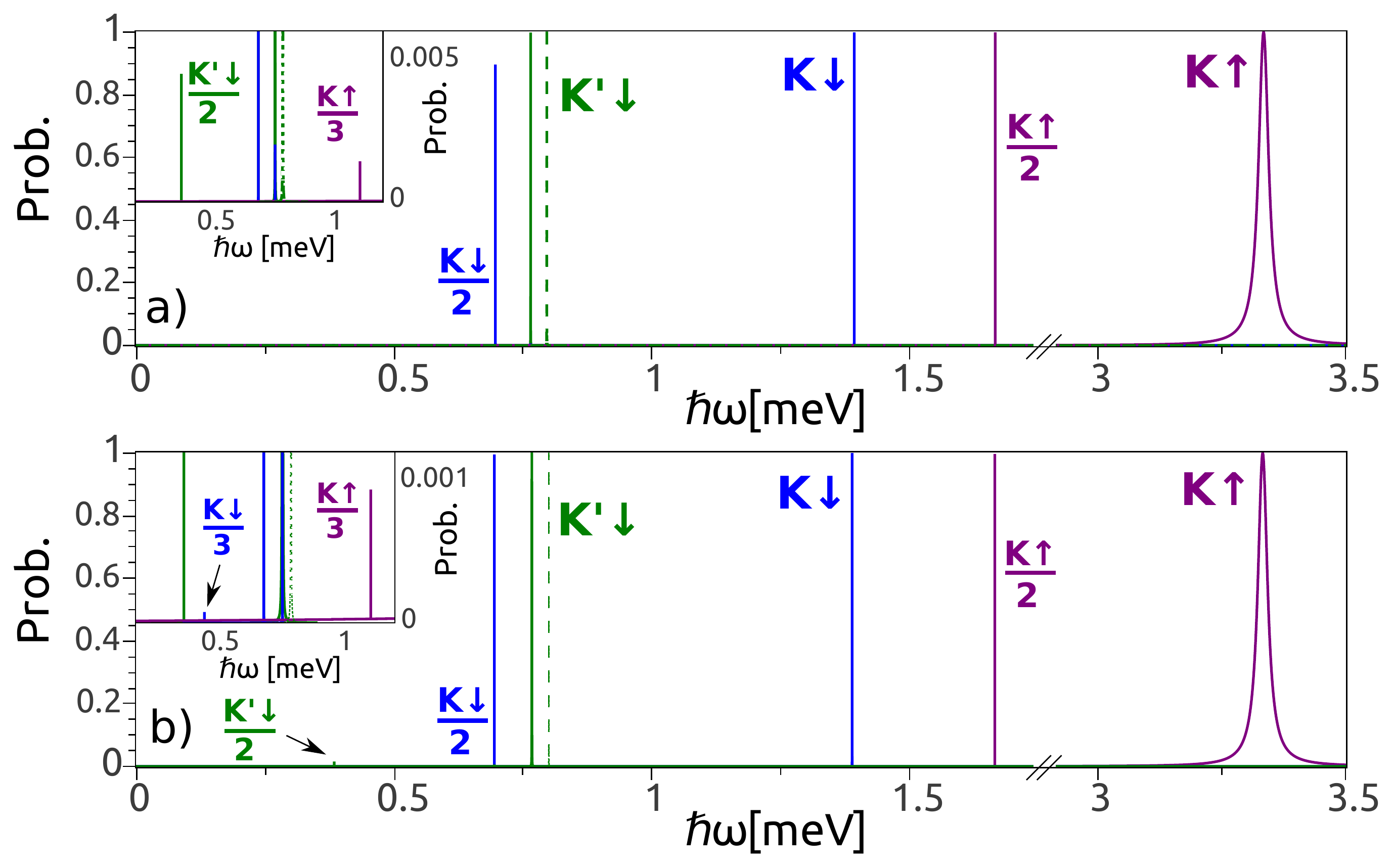}
\caption{Same as Fig. \ref{ss} only with SO coupling for a straight (a) or bent (b) CNT. The inset shows the half resonances at the lower energy side.
Solid (dashed) lines indicate the transitions with (without) a vacancy. In (a) the electric field in the $x$ direction is present. }
\label{szkanb}
\end{figure}

\subsection{Bent CNT}
The scan of the transitions in the evolution lasting 500 ns for the system corresponding to Fig. \ref{szkant} [straight CNT with a vacancy and $F_x=100$ kV/cm] is displayed in Fig. \ref{szkanb}(a).
When the disorder is present within the lattice the electric field perpendicular to the axis of the CNT only shortens the transition times [see Fig. \ref{time}] and does not open any new transition channels.
The perpendicular electric field can be replaced by the bend of the disordered nanotube with no difference for the transitions -- see the results of Fig. \ref{szkanb}(b) (scan with the bend, a vacancy present and $F_x=0$).
When the disorder is removed we obtain only the direct spin-flip transition with preserved valley and its fractional resonances -- see
both panels of Fig. \ref{szkanb} for the dashed lines obtained the spin-flip direct and fractional transitions with preserved valley.

For perfectly ordered lattice and straight CNT, without the component of the electric field perpendicular to the tube axis {\it we do not}
find {\it any} transitions in the considered energy range in spite of the presence of the SO coupling.

\begin{table*}[htbp]
\begin{tabular}{|c|c|c|c|}
\hline
 & {$|K'\uparrow\rangle \rightarrow |K'\downarrow\rangle$} & {$|K'\uparrow\rangle\rightarrow  |K\downarrow\rangle$} & {$ |K'\uparrow\rangle\rightarrow |K\uparrow\rangle$}\\ \hline \hline
straight CNT no vacancy $F_x=0$ & $\infty$ & $\infty$ & $\infty$ \\ \hline
straight CNT no vacancy $F_x=100$ kV/cm & 31 ns & $\infty$ & $\infty$ \\ \hline
straight CNT with vacancy $F_x=0$ &  48 ns & 342 ns & 167 ps \\ \hline
straight CNT with vacancy $F_x=100$ kV/cm & 23 ns & 102 ns & 167 ps\\ \hline \hline
bent CNT no vacancy $F_x=0$ & 42 ns & $\infty$ & $\infty$ \\ \hline
bent CNT with vacancy $F_x=0$ & 24 ns & 114 ns & 176 ps  \\ \hline \hline
& {$|K'\downarrow\rangle\rightarrow |K'\uparrow\rangle$} & {$|K'\downarrow\rangle\rightarrow  |K\uparrow\rangle$} & {$ |K'\downarrow\rangle\rightarrow |K \downarrow\rangle$} \\ \hline
straight CNT with vacancy, no SO & $\infty$ & $\infty$  & 168 ps \\ \hline
\end{tabular}
\caption{Direct transition times from the lowest-energy QD-localized level at 5 T with spin flip (second column), spin-valley transition (third column)
and valley transition (last column).  Spin-orbit coupling is included in all the cases but the last row.
In each a different driving frequency $\omega$ of AC field tuned to resonance was applied. The amplitude of AC field along the $z$ axis was set to 4 kV/cm. }
\end{table*}

\subsection{Transition times and selection rules for the direct transitions}

Summary of the direct transition times is given in Table I with spin-flip in the second column, spin-valley transition in the third column,
and valley transition with conserved spin in the last column.
The SO coupling due to the curvature of the nanotube is not enough to allow for spin transitions induced by AC field.
The spin transitions do appear with {\it (i)} a disorder {\it (ii)} the electric field perpendicular to the axis or {\it (iii)} the bend of CNT
in the segment with the QD. The presence of the disorder alone for a CNT with SO coupling is enough to drive all the three types
of transitions. Without disorder the bend and the perpendicular electric field only drive the spin flips and not the inter-valley transitions.
The inter-valley transition when allowed by the presence of the disorder is much faster than transitions including spin-flips.
The spin-flips (second column) that are present without disorder appear much faster when the disorder is introduced.

\begin{figure}[htbp]
\includegraphics[scale=0.21]{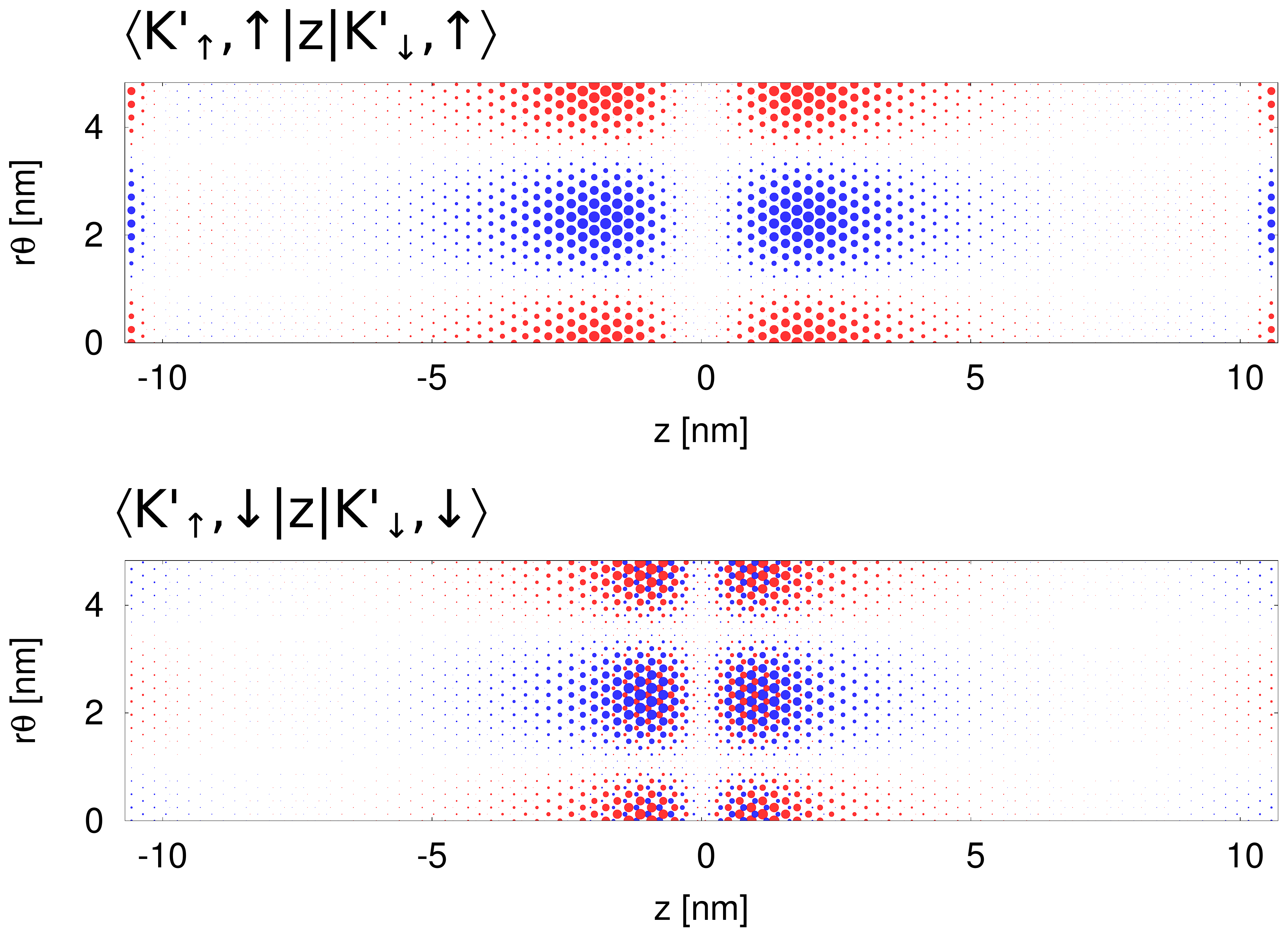}
\caption{The real part of spin-up and spin-down components of the integrand for $\langle K'\uparrow|z|K'\downarrow\rangle$
matrix element. In the upper (lower) panel we plot the product of the majority (minority) spin component for the initial state and minority (majority) spin component of the finial state.
The blue / red dots correspond to positive (negative) values.
}
\label{pd}
\end{figure}

\begin{figure}[htbp]
\includegraphics[scale=0.4]{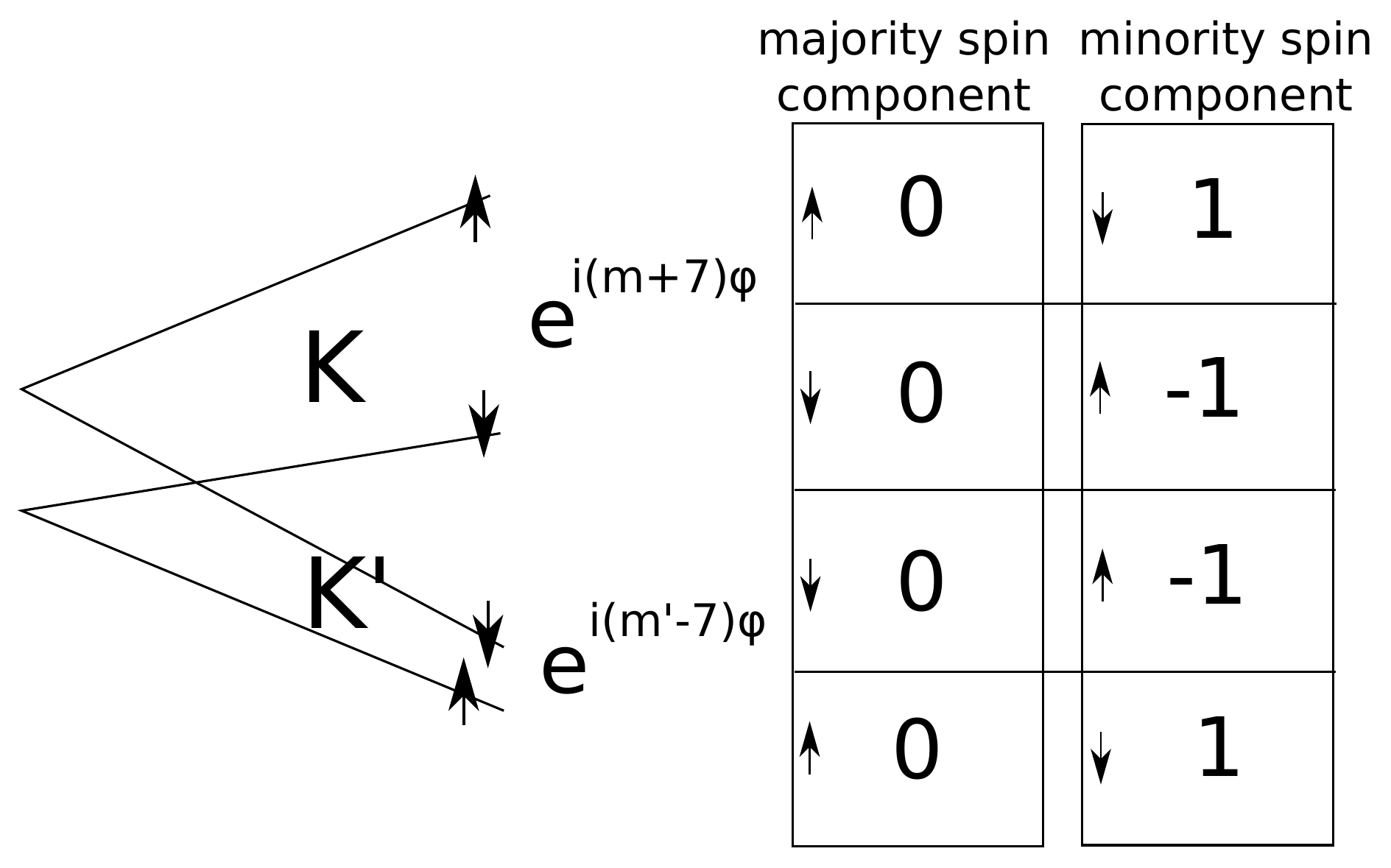}
\caption{The angular quantum numbers $m$ and $m'$ for the majority and minority spin components for a straight and clean CNT.
The numbers are given according to the order of energy levels at the right end of the schematically drawn spectrum.}
\label{mnie}
\end{figure}

\begin{figure}[htbp]
\includegraphics[scale=0.35]{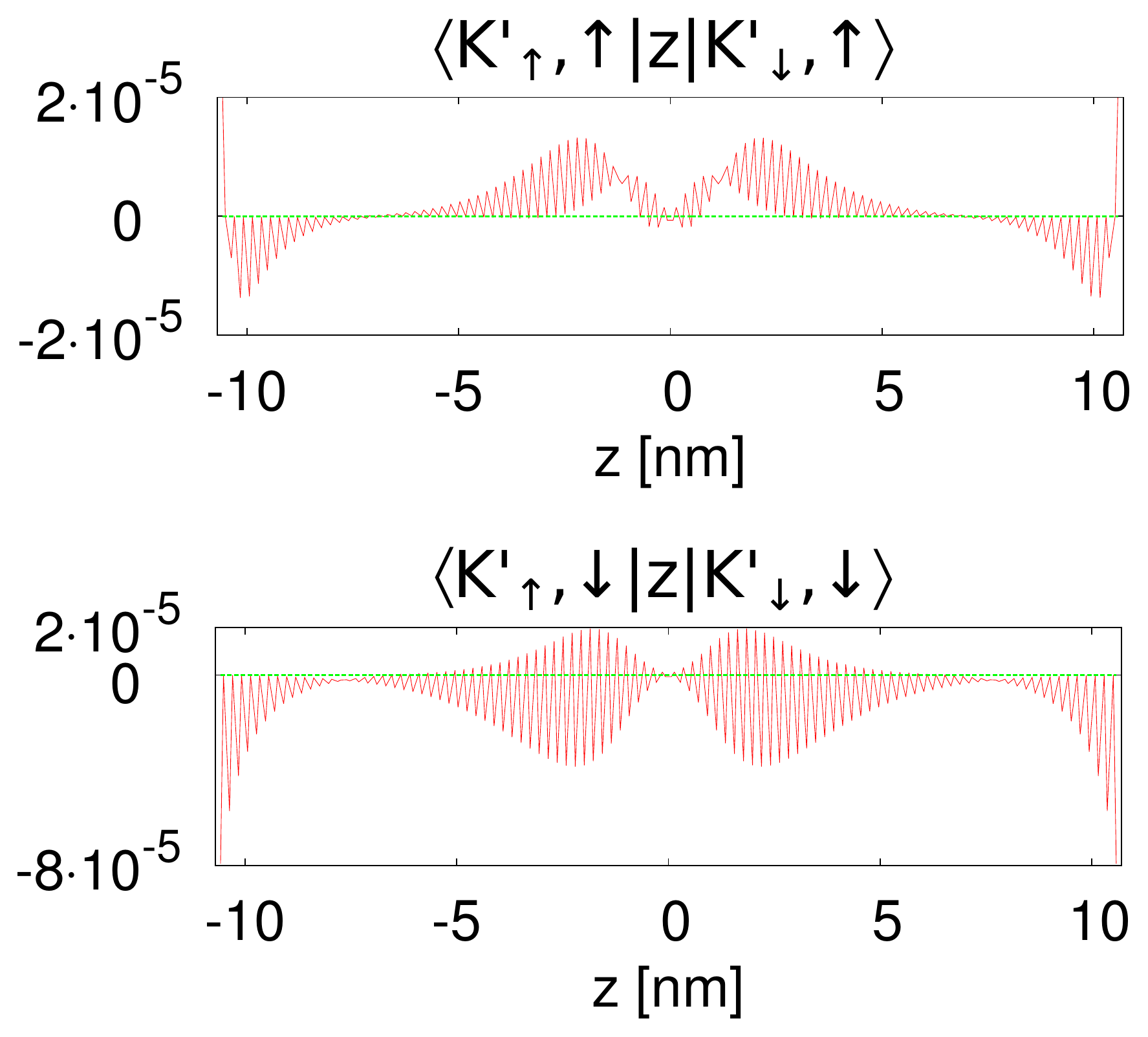}
\caption{Same as Fig. \ref{pd} only integrated over the angle $\theta$.
The green (red) curve correspond to the straight and clean CNT and $F_x=0$ ($F_x=100$ kV/cm).}
\label{zppz}
\end{figure}

The time dependence of Fig. \ref{time} for resonant frequencies indicate a pure two-level Rabi oscillation.
In agreement with the Rabi mechanism we find that the transition times of Table I are inversely proportional to the absolute value of the matrix element $\zeta_{if}=\langle i|z|f\rangle$,
with $i$ and $f$ standing for the initial and final states.
This matrix element also determines the selection rules of the discussed transitions.
Let us consider the straight and clean nanotube, for which no spin-flip transition is observed in spite of the presence of the SO coupling.
The spin-up and spin-down components of the integrands of the matrix element are given in Fig. \ref{pd}.
We find that the integrand function vanishes upon integration over the angle $\theta$.
The majority components of all the states considered in the context of the spin-valley transitions correspond to quantum number $m=m'=0$  (see the discussion of Fig. \ref{dm}).
The calculation of the matrix element for the spin-flip involves products of the majority components of the initial state with the minority components
of the final state, and vice versa. We extracted the quantum numbers for the minority spin components with the procedure explained in the context of Fig. \ref{dm}
with the results summarized schematically in Fig. \ref{mnie}.
The minority components differ by $\pm 1$ from the majority ones, hence the orthogonality over the angle which leads to the absence of the spin-flip
transitions for the straight and clean CNT. In order to allow for the transition that are forbidden by the angular symmetry to occur
one needs to lift the symmetry of the CNT along the circumference.
The electric field of about 100 kV/cm perpendicular to the axis or the bend of the nanowire allows for the spin-flip transition (see Fig. \ref{zppz}).
The bend of the CNT is effective in activating the transitions because of its interplay with the Gaussian potential forming the QD [$W_{QD}$] which
as due to external potential depends on $z$ only and ignores the deformation of the CNT.
In consequence at the bent part of the tube the atoms at the circumference perpendicular to the axis occupy locations with different potentials.
The selection rules are restored when the radius of the arc increases:
we found that the transition times for the spin-flip of a clean CNT scale linearly with $R$ [Fig. \ref{R}(a)].
On the other hand for straight CNT the spin-flip time scales as an inverse of the perpendicular electric field [Fig. \ref{R}(b)].

Note, that for both the straight and clean CNT the states of opposite valleys with $m=- m'$ remain orthogonal by the angular dependence.
In particular for $m=m'=0$ for the considered zig-zag CNT with 20 atoms along
the circumference the angular dependence is $\exp(i7\phi)$ and $\exp(-i7\phi)$ for both valleys.
In terms of the transition matrix elements this results in the absence of the inter-valley transition for a clean CNT.

\begin{figure}[htbp]
\includegraphics[scale=0.35]{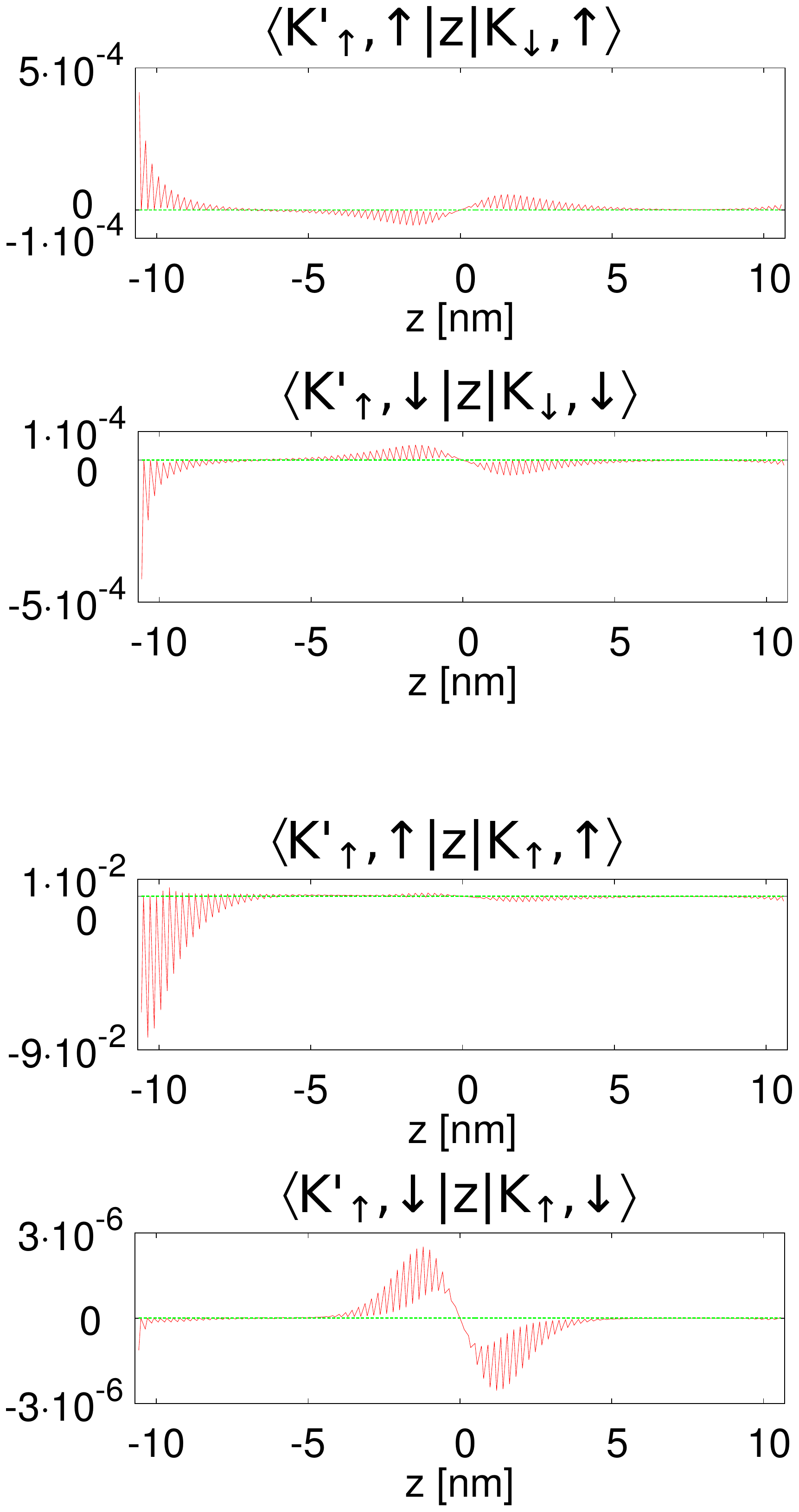}
\caption{Components of the integrands for the transition matrix element for a straight CNT with a vacancy near the left end of the tube [see Fig. 1(a)].
The green (red) curve correspond to the straight and clean CNT and $F_x=0$ ($F_x=100$ kV/cm).}
\label{dppd}
\end{figure}

\begin{figure}[htbp]
\begin{tabular}{ll}
a)\\
\includegraphics[scale=0.4]{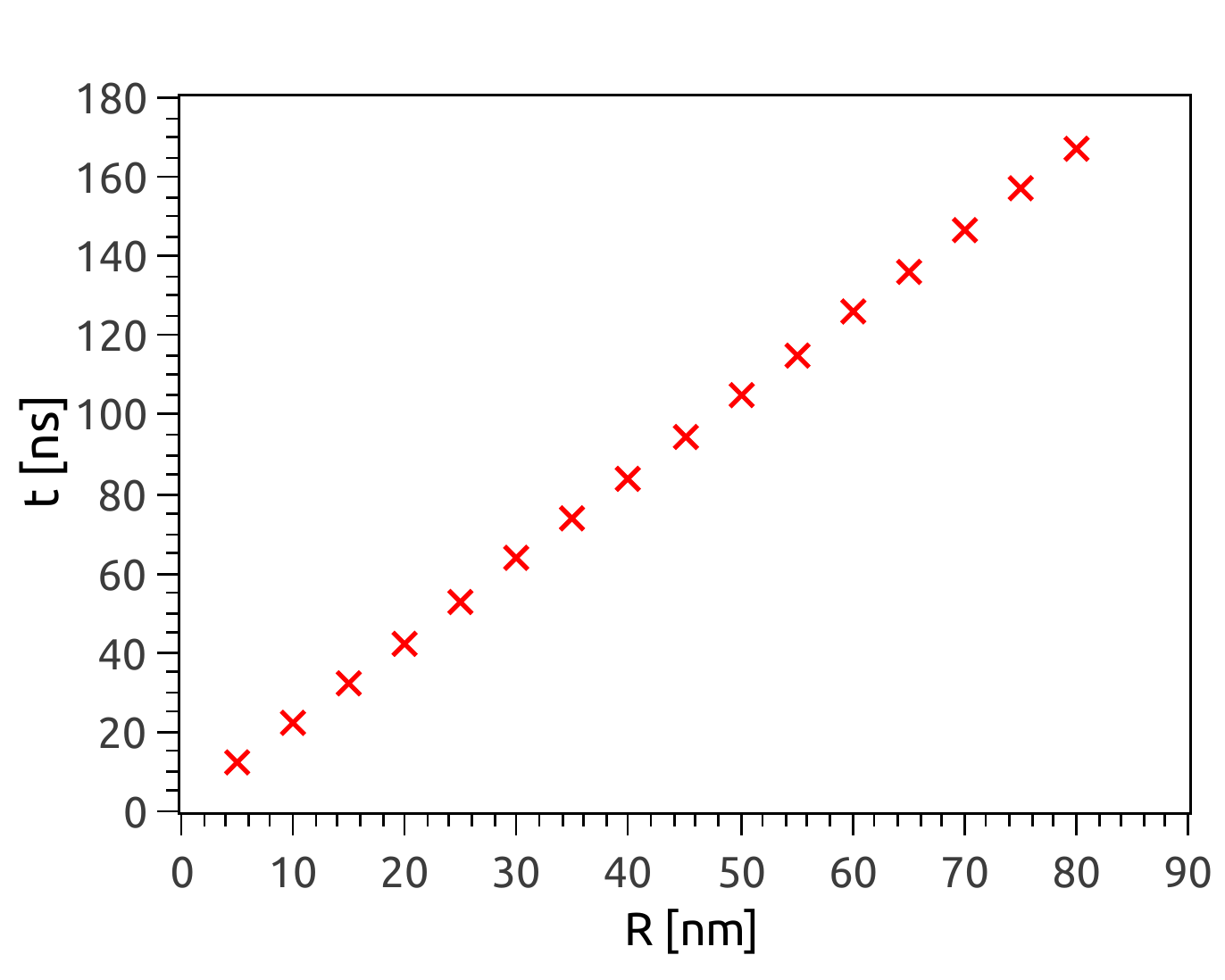} \\
b)\\
\includegraphics[scale=0.4]{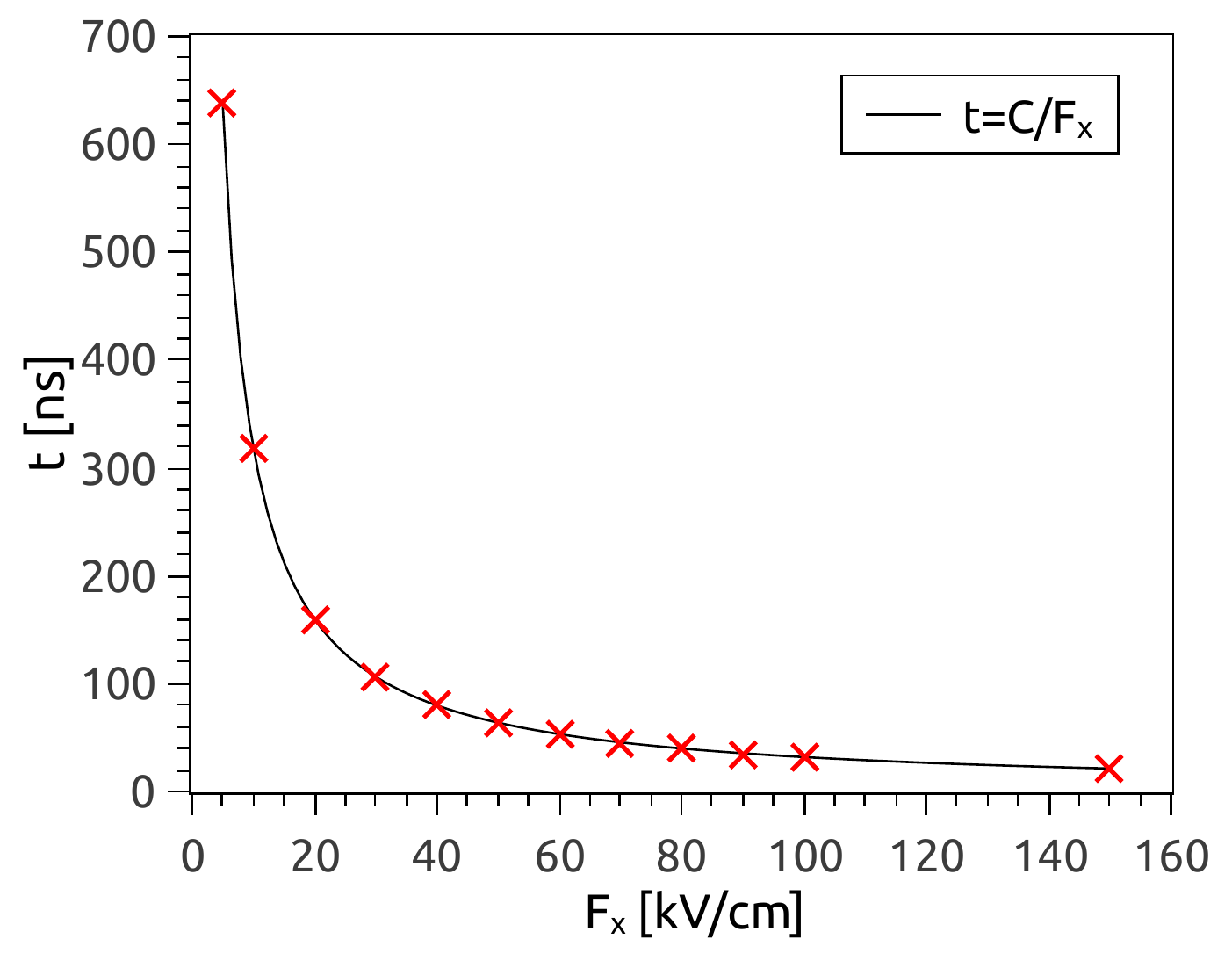} \\
c)\\
\includegraphics[scale=0.4]{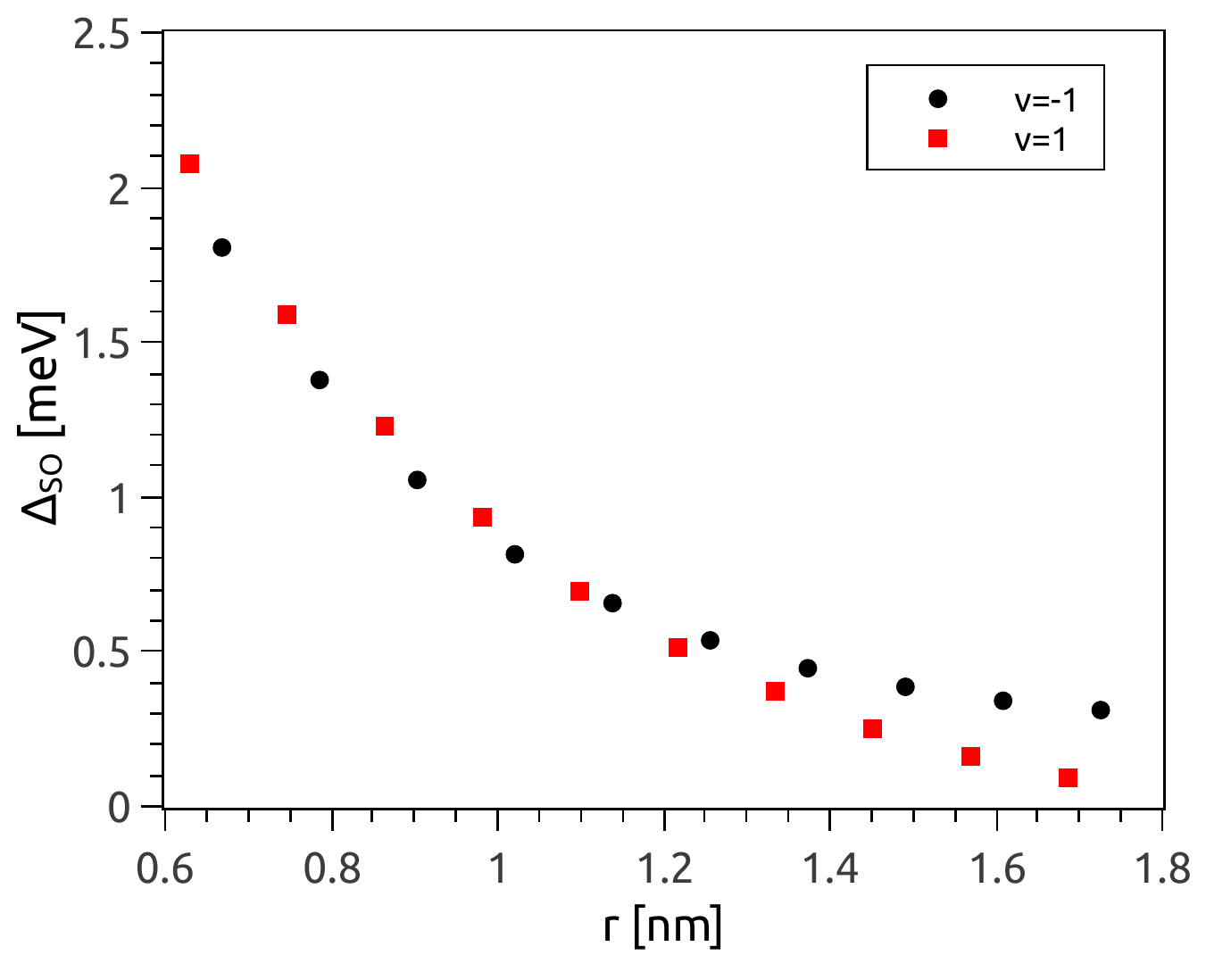} \\
d)\\
\includegraphics[scale=0.4]{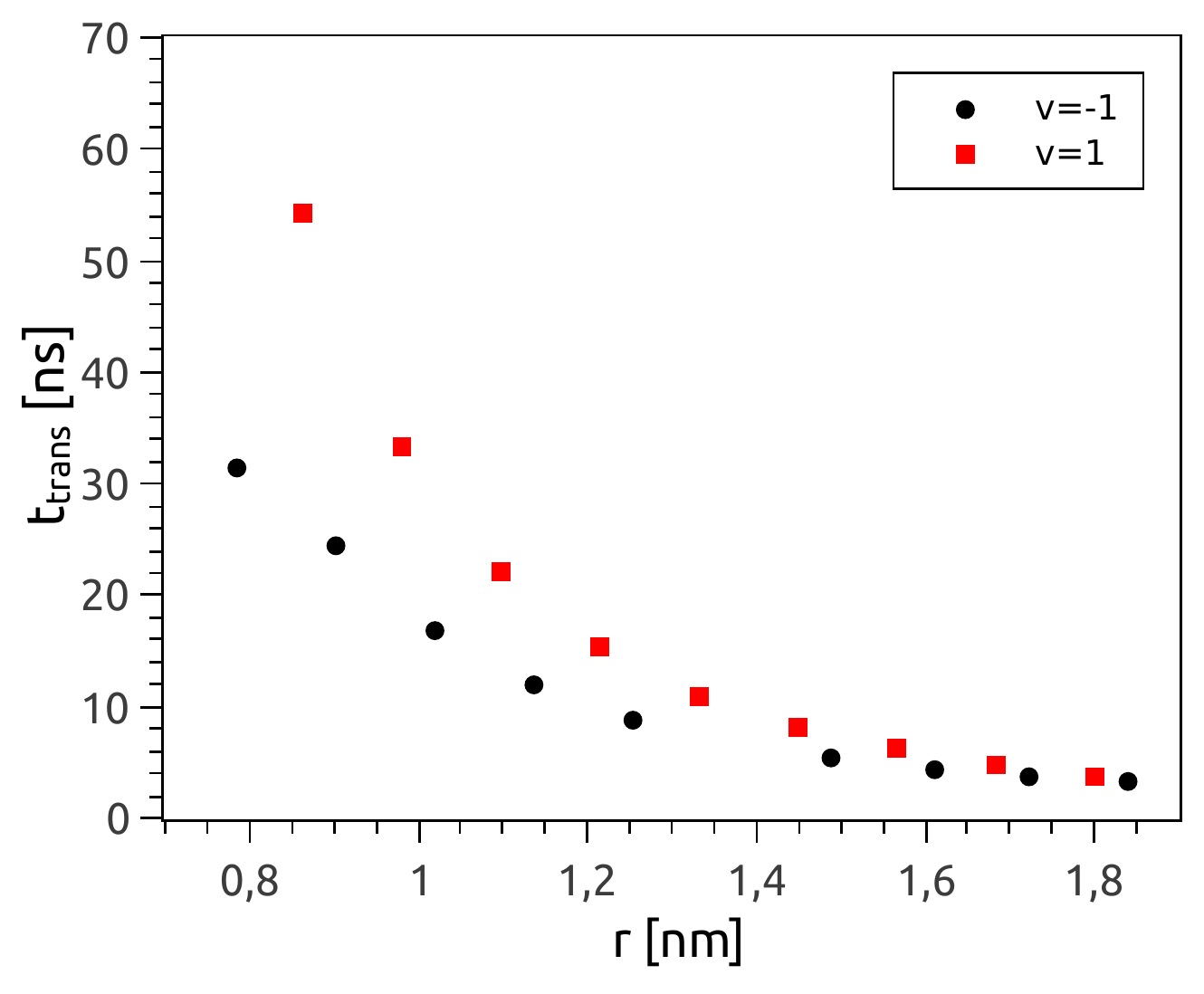} \end{tabular}
\caption{(a) Spin-flip transition $|K'\uparrow\rangle\rightarrow |K'\downarrow\rangle$ times as a function of the radius of the bent for a clean CNT
for $B=5$ T and AC amplitude of $4$ kV/cm for $r=0.78$ nm (as everywhere else in this paper) (b) Same as (a) only for a straight CNT in function of the electric
field perpendicular to the axis of the CNT. (c) SO energy, i.e. the splitting of the Kramers doublets at $B=0$, in function of the radius of the CNT for the straight and clean CNT without external fields.
(d) Same as (b) only in function of the radius of the CNT. Here the value of the perpendicular electric field was changed to keep the potential difference
at the opposite sides of the CNT constant i.e. $F_x=$ 100 kV/cm $\cdot\,(0.78$ nm $/r)$. }
\label{R}
\end{figure}

{
The disorder due to the vacancy near the end of the straight CNT 
not only introduces the asymmetry of the integrands with respect to the center of the QD [see Fig. \ref{dppd}]
but also perturbs the angular dependence of the wave function. Thus, the atomic disorder introduced by the vacancy introduces strongly non-zero transition elements 
between states of opposite valleys by angular symmetry, in a similar manner as the $F_x$ electric field does. However, the symmetry perturbation by the vacancy
is very strong as compared to the $F_x$ field of the order of 100 kV /cm. In presence of the disorder the spin-conserving intervalley transition is then very fast
and does not react on additional symmetry perturbation by the $F_x$ field,  c.f. $|K'\uparrow\rangle\rightarrow|K\uparrow\rangle$ transition times in Table I - see also Fig. 7.
}

For completeness in Fig. \ref{R}(d) we plotted the spin-flip transition times in function of the radius of the straight semiconducting CNTs.
In this plot the value of the perpendicular electric field is changed for different $r$ to keep the fixed potential difference at the opposite sides of the
CNT. Electric field applied to the CNT of radius $r$ is given by $F_x=$ 100 kV/cm $\cdot\,(0.78$ nm$/r)$. We found that the time scales exponentially with the radius of the CNT, in spite of the fact that SO coupling energy
decreases with $r$ see Fig. \ref{R}(c). The symmetry breaking effect of the perpendicular electric field is stronger for larger $r$ even
at a constant potential difference across the CNT. The fact that SO is reduced with $r$ does not prevent faster transition times for larger $r$. Note that the spin-transitions are forbidden
even for strong SO interaction unless the angular symmetry is broken.
In Fig. \ref{R}(c,d) we found that the results are organized in two series depending on the number of the atoms on the circumference
of the CNT with $N=3i+\nu$, with $i$ being an integer. The SO energy  scales as $1/r$ for $\nu=1$, for $\nu=-1$ at higher
values of $r$ the SO coupling energy vanishes slower.

\subsection{Fractional resonances}
\label{tdpt}
The direct solution of the Schr\"odinger equation produces both the direct transitions for $\omega^1_{AC}=\Delta E/\hbar$ and fractional resonances
for $\omega^n_{AC}=\Delta E/n\hbar$, which were clearly observed in the recent experiment of Ref. \cite{ffpei} [Fig. 4(e)].
The fractional resonances appear slower and possess finer widths [see Fig. 6(a) and Fig. 8].
We found that the fractional resonances can be explained within the time dependent perturbation theory.
Let us consider the second-order perturbation calculus to evaluate the expansion coefficients $c_n(t)$ of Eq. (7) as
\begin{equation} c_n(t)=c_n(0)+c_n^1(t)+ c_n^2(t). \label{tdpte}\end{equation}
With perturbation $V({\bf r},t)$, the matrix element
$V_{ni}=\langle \Psi_n |V({\bf r},t)|\Psi_i \rangle$, and $\omega_{ni}=(E_n-E_i)/\hbar$ the formulas for corrections read
\begin{equation}
c_{n}^{(1)}(t)=-\frac{i}{\hbar}\int_{t_{0}}^{t}dt'V_{ni}(t')e^{i\omega_{ni}t'},
\end{equation}
for the first order,
\begin{equation}
c_{n}^{(2)}(t)=-\frac{1}{\hbar^{2}}\sum_{m}\int_{t_{0}}^{t}dt'\int_{t_{0}}^{t'}dt"V_{nm}(t')V_{mi}(t")e^{i\omega_{nm}t'+i\omega_{mi}t"},
\end{equation}
for the second order.

In the time-dependence of the AC potential [Eq.(8)]  we have $\sin(\omega t)=\frac{1}{2i}\left(\exp(i\omega t)-\exp(-i\omega t)\right)$,
the first exponent produces very fast varying term when superposed on the transition frequency $\omega_{ni}$ which
do not drive any transitions in the considered range of the $\omega_{AC}$.
For that reason, we use the standard approach to set the perturbing potential to $V'=eF_0z (-\frac{1}{2i}\exp(-\omega t))$ which allows
for an analytic evaluation of corrections. Then we obtain
\begin{eqnarray}
c_{n}^{(1)}(t)&=&\frac{eF_0}{\hbar}\zeta_{ni} e^{i\frac{(\omega_{ni}-\omega)t}{2}}\frac{\sin(\frac{\omega_{ni}-\omega}{2}t)}{(\omega_{ni}-\omega)}
\end{eqnarray}
and
\begin{eqnarray}
c_{n}^{(2)}(t)&=&\frac{e^{2}F_0^{2}}{2i\hbar^{2}}\sum_{m}\zeta_{nm}\zeta_{mi}\nonumber \\&&\big[e^{i\frac{\omega_{nm}+\omega_{mi}-2\omega}{2}t}\frac{\sin(\frac{\omega_{nm}+\omega_{mi}-2\omega}{2}t)}{(\omega_{mi}-\omega)(\omega_{nm}+\omega_{mi}-2\omega)}\nonumber \\ && -e^{i\frac{\omega_{nm}-\omega}{2}t}\frac{\sin(\frac{\omega_{nm}-\omega}{2}t)}{(\omega_{mi}-\omega)(\omega_{nm}-\omega)}\big].
\end{eqnarray}

We use the expansion (\ref{tdpte}) to simulate the dynamics of the system instead of the direct solution to the Schr\"odinger equation.
The expansion does not conserve the norm which needs to be imposed for any $t$.
The results in the first order perturbation for the maximal occupation of the $K\downarrow$ level during $t=500$ ns evolution
for the initially occupied $K'\downarrow$ ground-state
are displayed in Fig. 6(b) with the purple dots.
We can see that the direct transition is exactly reproduced by the first order perturbation theory but the half resonance is missing.
The half-resonance is reproduced by the second order perturbation [orange line in Fig. 6(b)] which still overlooks the 1/3 resonance [see the inset to Fig. 6(b)], etc.
The half-resonance appears due to the presence of  $(\omega_{nm}+\omega_{mi}-2\omega)=(\omega_{ni}-2\omega)$  in the denominator of the first term under the sum of Eq. (14).
Concluding, the half resonances appear with participation of the intermediate ($m$) energy levels in a transition sequence $i\rightarrow m\rightarrow n$,
for the transition frequency $\omega=\omega_{ni}/2$ that is independent of $m$.
The expression for the second-order perturbation has a zero limit for $\omega\rightarrow \omega_{mi}$.
Hence, the intermediate level $m$ opens a channel for higher order transitions although no extra accumulation
of the wave function in the intermediate level is seen.
 The large variation of the width of the resonances between the direct and fractional resonances are reproduced by the time dependent perturbation theory.
From the derived formulas it is evident that higher order transitions are activated for larger amplitude of AC field ($F_0$).

\begin{figure}[htbp]
\includegraphics[scale=0.32]{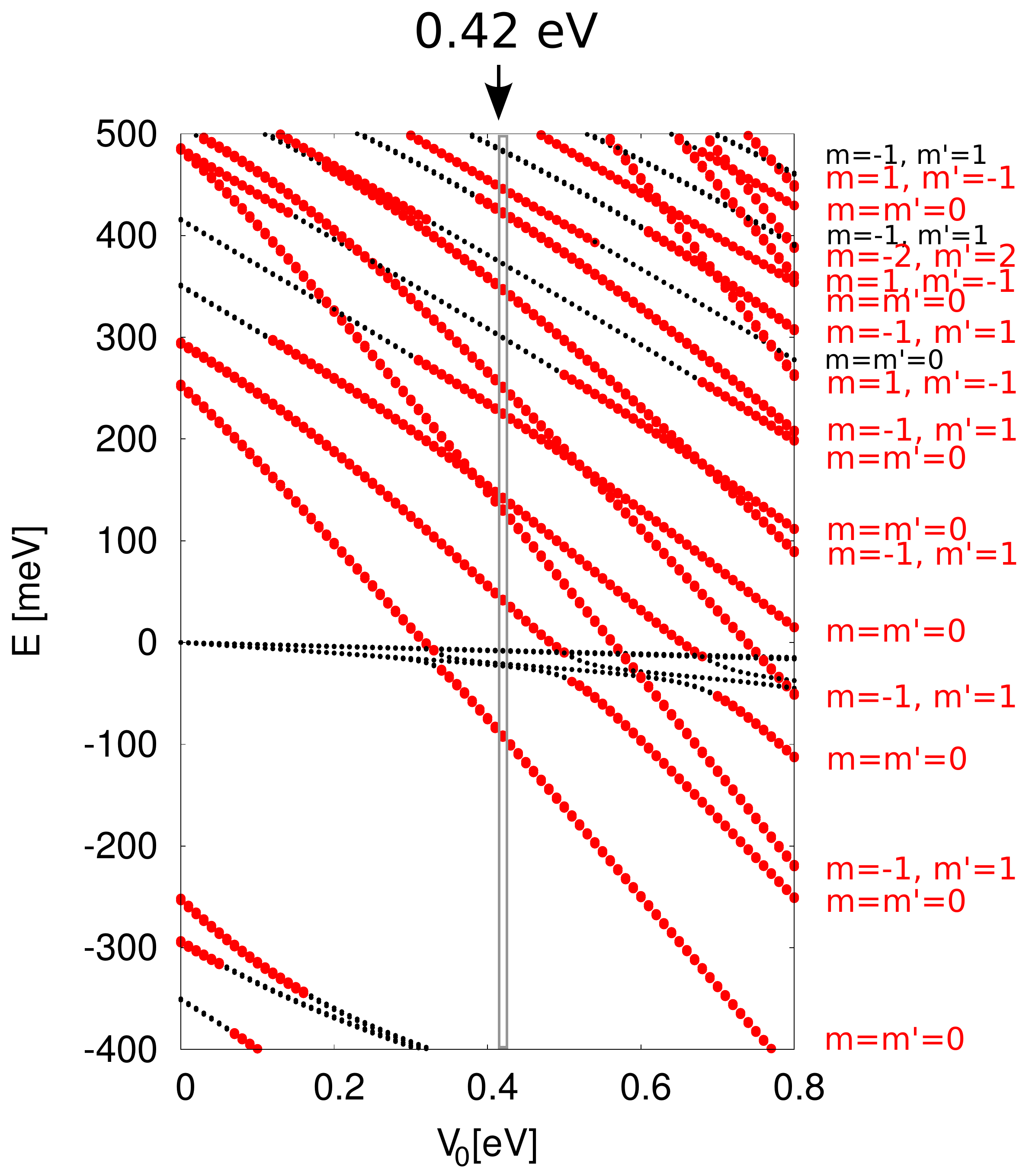} 
\caption{ Energy spectrum for the length of the QD increased from $2d=4.2$ nm [see Fig. 1(b)] to $2d=10.6$ nm. With the red dots
we marked the energy levels with 50\% of the probability density within the QD area ($[-d,d]$).
On the right hand side of the figure the quantum numbers $m$ ($m'$) defining variation of the wave functions along the circumference
of the nanotube for the states of valley $K$ ($K'$).
}\label{longerw}
\end{figure}

\begin{figure}[htbp]
\includegraphics[scale=0.42]{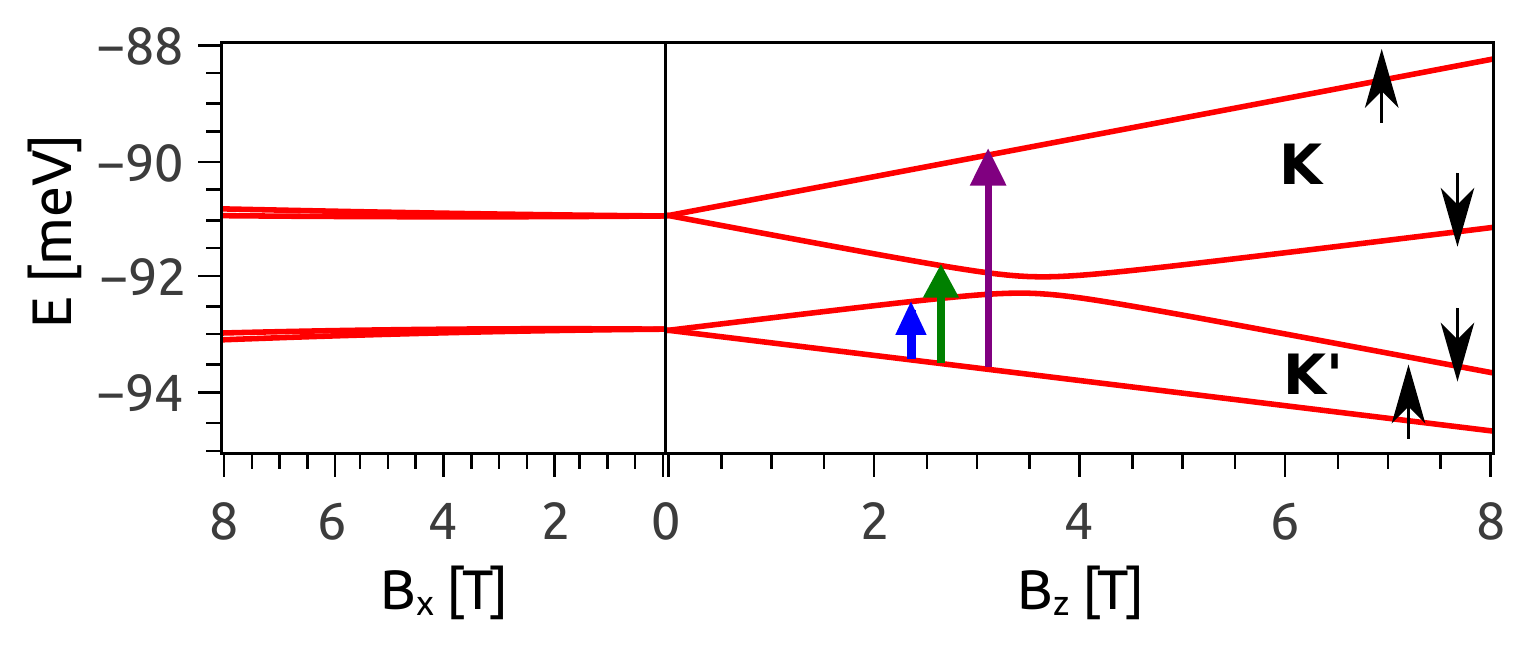}
\caption{ Energy spectrum for the lowest-energy QD localized electron energy level for $V_0=0.42$ eV as a function of the
external magnetic field oriented along the $x$ or $z$ axis for $2d=10.6$ nm. The plot is to be compared with Fig. 5(c) for $2d=4.2$ nm. 
The same SO coupling parameter is $\delta=0.003$, straight CNT of length $L=21.16$ nm is considered as in Fig. 5(c), position of the vacancy is kept unchanged [see Fig. 1(a)]. 
}\label{kritter}
\end{figure}

\begin{figure}[htbp]
\includegraphics[scale=0.42]{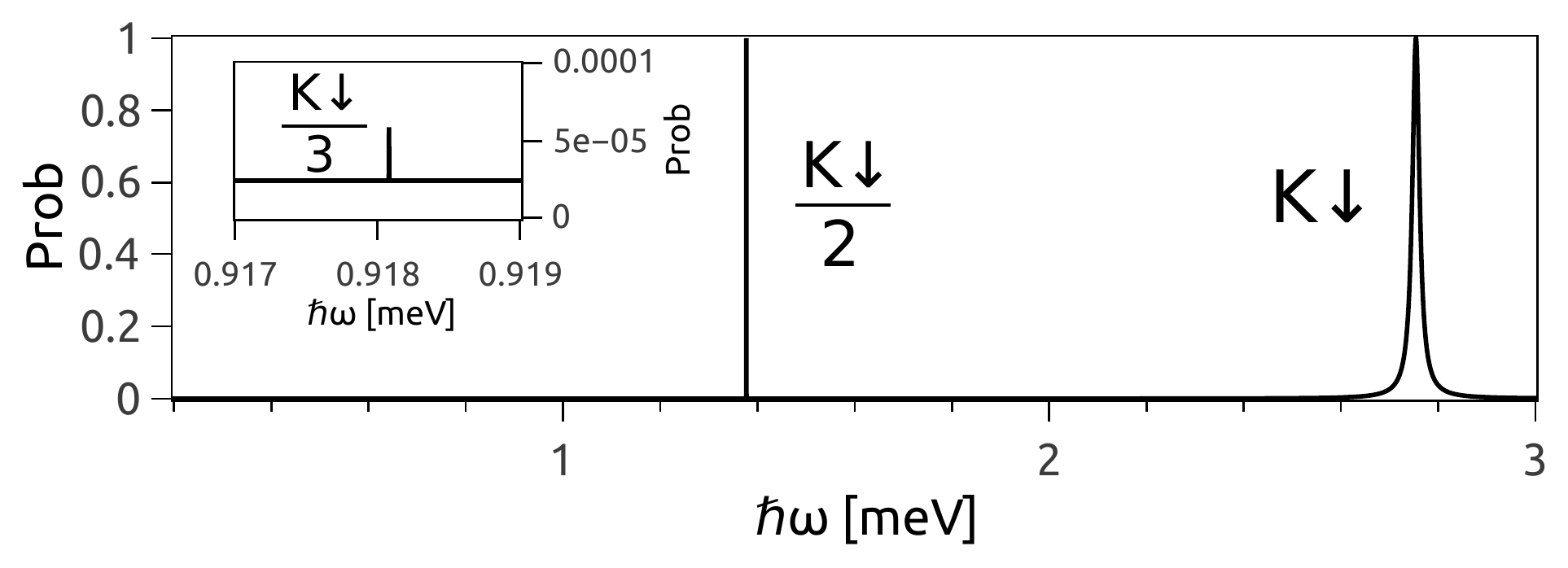}
\caption{
The maximal occupation of the $K  \downarrow$ energy level for simulations lasting 500 ns starting from
the $K'\downarrow$ ground-state as a function of the AC driving frequency $\omega$ for 
a straight CNT without SO coupling but with a vacancy near the left edge of the CNT [see Fig. 1(a)] in the carbon lattice and $F_{AC}=4$ kV/cm.
The wide and narrow peaks correspond to the direct ($K\downarrow$) and half-resonances ($K\downarrow/2)$.
The inset shows the zoom for the 1/3 of the excitation energy. The result -- obtained for $2d=10.6$ nm is to be compared 
with Fig. 6(a) presented above.
}\label{szkanlo}
\end{figure}

{
\subsection{Longer QD}
The present atomistic approach sets limits
to the size of the system that can be effectively studied. In the experiments, the QDs defined within CNT are by at least an order of magnitude longer than the radius of the tube (here $0.78$ nm).
However, the length of the QD does not seem to be a crucial parameter for the discussed spin and valley dynamics since the discussed transitions 
occur for a quadruple energy levels with varied spin and valley but nearly the same spatial charge distribution along the CNT. 
Nevertheless, in order to verify how robust the presented spin and valley dynamics picture is against the geometrical parameters of the quantum dot potential, we performed
calculations for the length parameter of the CNT increased from $2d=4.2$ nm to $2d=10.6$ nm. The other parameters, the length of the entire CNT ($L$) in particular are kept the same.

In Figure \ref{longerw} we plotted the energy spectrum in function of the QD depth. For $V_0=0$ the results of Fig. 1(b) for shorter QD are reproduced. 
As compared to the shorter QD we notice for $V>0$, that {\it i)} more energy levels fits into the longer QD as compared to $2d=4.2$ nm; 
{\it ii)} the coupling of the zigzag CNT edges to the dot localized states is stronger near the zero energy; and {\it iii)} the excited energy levels for fixed $m,m'$ go distinctly down
the energy scale. In particular, the first excited energy level has the same angular symmetry $m'=m=0$ as the ground-state so the lowest excitation becomes longitudinal instead
of the angular.

In the following we chose $V_0=0.42$ eV as the depth of the confinement potential, for which the magnetic field dependence of the lowest quadruple of dot-localized
energy levels is similar to the one discussed above [Fig. 5]. The $B$ dependence of the energy spectrum is given in Fig. \ref{kritter} for
the vacancy still at $l_v$ from the left edge of the CNT. 
The spin-orbit gap at $B=0$ is 1.95 meV as compared to 1.4 meV in Fig. 5; the width of the avoided crossing due to the intervalley transition is kept at $\simeq 0.28$ meV 
and the avoided crossing  occurs near $B_z=3.5$ T, as above.

Fig. \ref{szkanlo} shows the scan of the spin-conserving intervalley transitions obtained for a straight CNT without SO coupling, to be compared with Fig. 6(a) for the shorted QD.
Same parameters, including the amplitude of the AC field are applied. The same picture is obtained, with wide direct and thinner half-resonant transition.
The third order transition probability - for the simulation lasting 500 ns as above - is reduced significantly. Generally, we find that the transition times are longer
for the new set of parameters. The calculated values for the direct transitions are given in Table II. The transition times increase by a factor between 1.4 and 2.7, depending
on the transition type and specific case. Note, that the decrease of the spin and  valley transition rate is observed in spite of the fact that the energy variation 
within the CNT induced by the AC field is larger for a longer dot. In consistence with the data of Table II, we found a reduction of the transition matrix elements 
for both the spin and valley transitions. Two facts are responsible for this reduction: {\it i)} We find that the Hamiltonian eigenstates for the longer QD  are 
closer to the $\sigma_z$ eigenstates, i.e. the spins of the initial and final states are better defined for the longer QD. The effects of the spin-orbit coupling are 
naturally weaker for a larger confinement volume. {\it ii)} The lowest-energy confined level of the longer QD shifts down from about -40 meV to about -92 meV, which 
induces weaker coupling of the QD states with the zigzag edge states -- of energy $\simeq 0$. In the described system the vacancy is localized close to the left edge of
the CNT at a distance of $l_v$ from the end of the tube, so reduction of the QD-edge coupling is accompanied by reduction of the disorder induced intervalley scattering. We found that the part of the probability density
localized within the distance of $2l_v$ from the left edge is reduced by 50\% for the longer dot as compared to the case discussed above.

Concluding, we find that the length of the CNT does not have a qualitative effect on the spin and valley dynamics in the QD driven by AC field,
at least in the discussed case of transitions within the quadruple of energy levels of same spatial distribution split only by the spin-orbit and
intervalley coupling.

\begin{table*}[htbp]
\begin{tabular}{|c|c|c|c|}
\hline
 & {$|K'\uparrow\rangle \rightarrow |K'\downarrow\rangle$} & {$|K'\uparrow\rangle\rightarrow  |K\downarrow\rangle$} & {$ |K'\uparrow\rangle\rightarrow |K\uparrow\rangle$}\\ \hline \hline
straight CNT no vacancy $F_x=0$ & { $\infty$} & { $\infty$ } & { $\infty$ } \\ \hline
straight CNT no vacancy $F_x=100$ kV/cm & { 85 ns } & { $\infty$ } & { $\infty$ } \\ \hline
straight CNT with vacancy $F_x=0$ &   { 67 ns } & { 470 ns } & { 231 ps } \\ \hline
straight CNT with vacancy $F_x=100$ kV/cm & {  42 ns } & {  206 ns } &  {  231 ps } \\ \hline \hline
bent CNT no vacancy $F_x=0$ & {100 ns } & {$\infty$ } & {$\infty$ } \\ \hline
bent CNT with vacancy $F_x=0$ & {  40 ns } & {  270 ns } & {  210 ps } \\ \hline \hline
& {$|K'\downarrow\rangle\rightarrow |K'\uparrow\rangle$} & {$|K'\downarrow\rangle\rightarrow  |K\uparrow\rangle$} & {$ |K'\downarrow\rangle\rightarrow |K \downarrow\rangle$} \\ \hline
straight CNT with vacancy, no SO & {  $\infty$ } & {  $\infty$ }  & {  231} ps \\ \hline
\end{tabular}
\caption{Direct transition times from the lowest-energy QD-localized level at 5 T with spin flip (second column), spin-valley transition (third column)
and valley transition (last column).  Spin-orbit coupling is included in all the cases but the last row.
In each a different driving frequency $\omega$ of AC field tuned to resonance was applied. The amplitude of AC field along the $z$ axis was set to 4 kV/cm.
The length of the QD is $2d=10.6$ nm. The values for $2d=4.2$ nm were given in Table I.}
\end{table*}
}

\section{Summary and conclusions}

In summary, we presented simulations of the spin flip and inter-valley transitions in a quantum dot defined
within a semiconducting carbon nanotube. We considered a single excess electron in the quantum dot and evaluated
the dynamics of the spin and valley transitions driven by external AC electric field.
Time-dependent calculations used the basis of localized eigenstates as determined by the tight-binding approach.

For a straight and clean CNT the spin-flips are forbidden even for strong SO coupling.
The spin transitions are triggered by electric field perpendicular to the axis of the CNT.
We demonstrated that the spin-flip transition times are inversely proportional
to the value of the perpendicular electric field component. We demonstrated that the bend of the CNT in external electric field allows for the spin-flips due to lifting of the selection
rules by lowering the angular symmetry of the eigenstates with the spin-flip transition times scaling linearly with $R$.

We demonstrated that when SO coupling is present the atomic disorder alone allows for all types of transitions including spin flips.
We discussed the disorder introduced by a vacancy which -- even when far from the QD -- perturbs the angular
symmetry of the eigenstates lifting the selection rules prohibiting the inter-valley transitions.
The inter-valley transitions when allowed by the lattice disorder appear roughly 100 to 1000 times faster than the spin flips
and are insensitive to the electric fields perpendicular to the axis of the CNT.

\section*{Acknowledgements}
This work was supported by National Science Centre
according to decision DEC-2013/11/B/ST3/03837, by PL-GRID infrastructure and by Ministry of Science and Higher Education
within statutory tasks of the Faculty. Calculations were performed
in ACK -- CYFRONET -- AGH on the RackServer Zeus.

\end{document}